\documentclass[conference,compsoc]{IEEEtran}
\ifCLASSOPTIONcompsoc
  \usepackage[nocompress]{cite}
\else
  \usepackage{cite}
\fi
\usepackage{amsmath}
\usepackage{nicefrac}
\usepackage{siunitx}
\usepackage{array,framed}
\usepackage{booktabs}
\usepackage{makecell}
\usepackage{amssymb}
\usepackage{pifont}

\usepackage{
  color,
  float,
  epsfig,
  wrapfig,
  graphics,
  graphicx,
  subcaption
}
\usepackage[table]{xcolor} 
\usepackage{textcomp,amssymb}
\usepackage{setspace}
\usepackage{latexsym,fancyhdr,url}
\usepackage{enumerate}
\usepackage[linesnumbered,ruled,vlined]{algorithm2e}
\usepackage{algpseudocode}
\usepackage{booktabs}
\usepackage{pifont}
\usepackage{booktabs}
\usepackage{multirow}
\newcommand{\cmark}{\ding{51}}
\newcommand{\xmark}{\ding{55}}
\usepackage{diagbox}  
\usepackage{threeparttable}
\usepackage{xparse} 
\usepackage{xspace}
\usepackage{multirow}
\usepackage{csvsimple}
\usepackage{balance}
\usepackage{enumitem}

\usepackage{pifont}
\usepackage[normalem]{ulem}
\usepackage{tikz}

\usepackage{
  tikz,
  pgfplots,
  pgfplotstable
}
\usepackage{hyperref}

\usetikzlibrary{
  shapes.geometric,
  arrows,
  external,
  pgfplots.groupplots,
  matrix
}

\usepackage{tikz} 

\newcommand{\copyrighttext}{%
  \footnotesize
  \textcopyright~2026 IEEE.
  Personal use of this material is permitted.
  Permission from IEEE must be obtained for all other uses,
  in any current or future media, including
  reprinting/republishing this material for advertising or
  promotional purposes, creating new collective works,
  for resale or redistribution to servers or lists,
  or reuse of any copyrighted component of this work
  in other works.
}

\newcommand{\copyrightnotice}{%
  \begin{tikzpicture}[remember picture,overlay]
    \node[anchor=south,yshift=10pt]
      at (current page.south) {%
        \fbox{%
          \parbox{\dimexpr\textwidth-2\fboxsep-2\fboxrule\relax}{%
            \copyrighttext
          }%
        }%
      };
  \end{tikzpicture}%
}

\begin{document}

\title{
{\normalfont\normalsize
Accepted by the 48th IEEE Symposium on Security and Privacy (IEEE S\&P 2027)
\par}
\vspace{0.4em}

ScrambleMark: Zero-Query Watermark-Agnostic Black-Box Evasion of Modern Speech Watermarking}
\author{

\IEEEauthorblockN{
Binhao Ma\IEEEauthorrefmark{4},
Weikang Ding\IEEEauthorrefmark{4},
Hanqing Guo\IEEEauthorrefmark{2},
Wenwei Zhao\IEEEauthorrefmark{3},
Xiaowen Li\IEEEauthorrefmark{3}
\\
Leah Ding\IEEEauthorrefmark{1},
Zhuo Lu\IEEEauthorrefmark{3},
Rui Duan\IEEEauthorrefmark{4}
}

\IEEEauthorblockA{
\begin{tabular}{cc}
\IEEEauthorrefmark{4}University of Missouri--Kansas City
&
\IEEEauthorrefmark{2}Indiana University Bloomington
\\[0.3em]
\IEEEauthorrefmark{3}University of South Florida
&
\IEEEauthorrefmark{1}American University
\end{tabular}
}
}

\maketitle

\copyrightnotice
\begin{abstract}

Speech watermarking is becoming a key defense for digital copyright and content authentication. Existing attacks treat evasion as a removal problem: they erase the watermark, relying on detector access or on speech data that is hard to collect, such as paired clean–watermarked or clean target speaker speech. Commercial platforms are moving to close off these assumptions. TrustedAudio, for instance, restricts detection to authenticated users and denies meaningful detection feedback to unauthorized parties. Adversaries thus cannot tell whether a sample is watermarked or which watermark it carries, nor reliably remove it without degrading speech quality.

In this work, we show that evading a speech watermark does not require erasing it: detection relies on the structural consistency of the embedded pattern, and disrupting this consistency is enough to defeat it. Building on this observation, we identify harmonic structure as an effective attack surface, as perturbing it breaks watermark consistency while preserving perceptual quality. We therefore propose ScrambleMark, a zero-query, watermark-agnostic, black-box attack that evades detection through harmonic scrambling. ScrambleMark combines a deterministic engine for discovering transferable harmonic perturbations with a generative model that learns randomized scrambling patterns, avoiding reliance on fixed attack rules. Across three commercial platforms (Google SynthID, Resemble AI, and TrustedAudio), ScrambleMark is the only attack to both evade detection and keep speech natural, achieving nearly 97\% evasion and a 4.2/5 human rating, far above the 1.1 to 1.7 of prior attacks. Beyond evasion, ScrambleMark withstands adaptive recovery designed to reverse it, evidence that structural evasion is hard to mitigate in practice.

\end{abstract}


%
\IEEEpeerreviewmaketitle

\section{Introduction}

Driven by recent advances in deep neural networks (DNNs)~\cite{su2018snr,zhao2022ssvs,chen2023wavmark,liu2023detecting,roman2024proactive}, audio watermarking has become a widely adopted technique for establishing provenance and protecting copyright in digital media~\cite{uddin2024review,kumsawat2008multiwavelet,hua2016twenty}. Among its applications, speech has emerged as a particularly important domain due to the rapid progress of voice cloning and speech synthesis systems, which enable adversaries to impersonate individuals~\cite{arup_cnn2024}, commit fraud~\cite{fbi_impersonation2025}, and spread disinformation~\cite{fcc_robocall2024}. To address these threats, modern speech watermarking systems embed imperceptible marks into audio that can later be verified to establish provenance and detect misuse~\cite{chen2023wavmark,triniMark2025,liu2023detecting,roman2024proactive,zongaudiomarknet}. However, despite these advances, watermarking systems remain vulnerable to evasion attacks~\cite{liu2024audiomarkbench,lopez2024speech}, in which an adversary manipulates a watermarked speech signal to render the embedded watermark undetectable. Such attacks undermine the reliability of watermark-based provenance and misuse detection.

The first wave of studies on watermark evasion attacks has shown that audio watermarking fails under a range of adversarial capabilities, each assuming a different form of access. With query access to the detector, an adversary can iteratively optimize noise perturbations that suppress the watermark signal until detection fails~\cite{liu2024audiomarkbench}. When the embedding process is exposed, the adversary can instead inject a secondary watermark that overwrites the original~\cite{yao2026yours}. Given knowledge of the watermark pattern, the watermark can be treated as structured noise and removed by a trained enhancement model~\cite{lopez2024speech}.

However, existing attacks are impractical in realistic deployment settings. Most rely on one or more strong assumptions: access to detector feedback~\cite{liu2024audiomarkbench,yao2026yours}, paired clean--watermarked data~\cite{lopez2024speech}, or clean target-speaker speech~\cite{yaoblack}. In practice, an adversary typically cannot tell whether a speech sample is watermarked, which scheme was used, or which detector verifies it, and so lacks the feedback and side information these attacks require. While an attacker could, theoretically, brute-force query a collection of public watermark detectors to guess which one applies, the commercial watermarking platform, TrustedAudio~\cite{trustedaudio_about}, is specifically designed to prevent such probing. Operationally, detection interfaces are account-scoped: only authenticated users can retrieve detection outcomes, whereas adversaries are denied informative feedback (e.g., receiving “not detected” responses) even when submitting speech that contains a legitimate user’s watermark. Therefore, access-dependent strategies become largely ineffective in the real world.


These constraints lead us to challenge an assumption shared by prior attacks: that successful evasion must erase the watermark. We therefore examine what modern watermark detectors actually rely on. To remain detectable under common distortions such as compression and noise, watermarking systems do not embed information at a single location; they spread it into a structured pattern across the time-frequency representation, and detection commonly works by matching this pattern. This exposes a new attack surface: because the watermark's robustness lies in this structural consistency, an adversary can defeat detection by disrupting that consistency rather than removing the watermark, so that the detector fails to match the pattern.

To bypass these real-world constraints, we propose ScrambleMark, a novel zero-query, watermark-agnostic, black-box attack that scrambles the harmonic structure of speech to disrupt the consistency required for watermark verification. Building such an attack under a strict black-box setting raises two challenges. (i) The attacker has no knowledge of where the watermark resides, and modern watermarking schemes often distribute watermark information across broad regions of the spectrum. Effective evasion therefore requires perturbations that generalize across the signal while preserving perceptual quality. (ii) The most straightforward way to disrupt watermark structure is through deterministic signal-processing transforms. However, fixed scrambling rules leave predictable perturbation patterns that an adaptive defender can estimate and partially reverse. Robust evasion therefore also requires randomized scrambling that avoids reliance on any fixed transformation.

To systematically resolve these challenges, ScrambleMark introduces a multi-stage synthesis framework. We first formalize the Evasion-Quality Trade-off Ratio (EQTR) across diverse perceptual feature domains, finding that harmonic-axis warping provides the best balance between disrupting watermark structure and preserving speech intelligibility. Building on this insight, we construct a surrogate-guided deterministic partition-shift engine that generates transferable, recovery-aware scrambling examples for training. Finally, to eliminate mathematical invertibility, we lift these deterministic operators into a conditional diffusion model governed by randomized target supervision and dynamic temporal scrambling. Through a pretraining and fine-tuning alignment scheme, ScrambleMark learns to confine its randomized perturbations exclusively to the harmonic dimension. This design lets the model generate natural speech on unseen watermarked speech while executing a recovery-resistant structural evasion attack.

We comprehensively evaluate ScrambleMark against seven state-of-the-art academic watermarking systems and three commercial platforms (Google SynthID, TrustedAudio, and Resemble AI). On commercial platforms, ScrambleMark achieves an average attack success rate (ASR) of nearly 97\% while maintaining a mean human rating of 4.2/5, substantially outperforming existing watermark removal attacks. Notably, even when the defender knows the attack algorithm, recovery reduces ScrambleMark's ASR by only 0.02. Audio samples for ScrambleMark and the baseline attacks are available on the
\href{https://scramblemark.github.io/}{project website}.

Our contributions are summarized as follows:
\begin{itemize}
    \item \textbf{Structural Evasion.}
    We introduce \textit{structural evasion}, a new attack paradigm that defeats robust watermarking schemes by disrupting watermark structural consistency under a strict zero-query, black-box setting. We design ScrambleMark, a conditional diffusion framework that implements this paradigm through randomized target supervision and dynamic temporal scrambling, avoiding fixed recovery paths for defenders.
    \item\textbf{Real-world Impact.} We present the first evasion study of commercial speech watermarking platforms, including Google SynthID, TrustedAudio, and Resemble AI. We further conduct human listening studies to evaluate perceptual quality and demonstrate the practical impact of ScrambleMark in real-world settings.
    \item \textbf{Defense Resilience.} We further evaluate ScrambleMark against two adaptive recovery defenses and show that it remains effective even when the defender knows the attack algorithm and has access to the original speech.
\end{itemize}

\section{Background and Design Intuition}
We first introduce speech watermarking background, then present our threat model, review existing evasion attacks, and describe the design intuition and system overview. \label{sec:background}

\begin{figure}[t]
    \centering
    \includegraphics[width=0.94\linewidth]{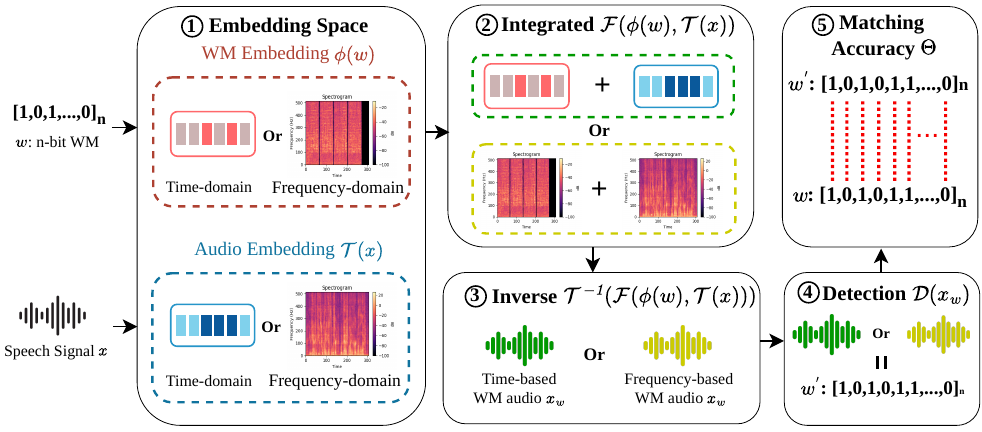}
    \caption{Overview of the watermarking pipeline.}
    \label{fig:overviewwm}
\end{figure}

\subsection{Speech Watermarking}\label{sec:watermark}
Speech watermarking is a key mechanism for copyright tracing, content attribution, and identity protection~\cite{hua2016twenty,uddin2024review}. While prior work has proposed academic systems such as AudioSeal~\cite{roman2024proactive} and Timbre~\cite{liu2023detecting}, commercial adoption is rapidly accelerating. Google SynthID~\cite{deepmind_synthid}, Resemble AI~\cite{resemble_watermarker}, and TrustedAudio~\cite{trustedaudio_about} have already deployed speech watermarking in real-world services.

\noindent\textbf{{Watermarking Pipeline.}}
As illustrated in Fig.~\ref{fig:overviewwm}, the watermarking framework is composed of two primary stages: watermark embedding (steps~\ding{172} to \ding{174}) and detection (steps~\ding{175} to \ding{176}).
The details of each step are described as follows:

\noindent\textbf{Watermarking Embedding.}  
The speech watermarking embedding takes two inputs: a speech signal \( x \) and an \( n \)-bit watermark \( w = \{w_i\}_{i=1}^n, \, w_i \in \{-1,1\} \)~\cite{liu2023detecting}. The embedding function $\mathcal{E}$ maps $(x,w)$ to the watermarked signal $x_w$ as
$x_w=\mathcal{E}(x,w)=\mathcal{T}^{-1}\!\left(\mathcal{F}(\mathcal{T}(x),\phi(w))\right)$.

\noindent\underline{\ding{172} Input transformation.}
The speech signal $x$ is mapped to an embedding space via $\mathcal{T}(x)$, chosen as either the identity function for time-domain embedding~\cite{manikandan2018histogram,hosny2018improving} or the Short-Time Fourier Transform (STFT) as a frequency-domain embedding~\cite{podilchuk2001digital,cox1996secure}. In parallel, the watermark vector $w$ is mapped by an embedding function $\phi(\cdot)$, such as a neural encoder~\cite{roman2024proactive}, into a structured pattern $\phi(w)$ that lies in the same domain as $\mathcal{T}(x)$ for subsequent integration~\cite{liu2023detecting}.

\noindent\underline{\ding{173} Integration.}  
The transformed audio \( \mathcal{T}(x) \) and the embedded watermark pattern \( \phi(w) \) are integrated via a fusion function \( \mathcal{F}(\cdot) \) to form a unified latent representation~\cite{chen2023wavmark,roman2024proactive}.

\noindent\underline{\ding{174} Inverse transformation.} Finally, the watermarked signal \( x_w \) is reconstructed by applying the inverse transformation \( \mathcal{T}^{-1}(\cdot) \), e.g., inverse STFT, recovering a time-domain signal.

\noindent\textbf{Watermarking Detection.}
The detector~\cite{liu2023detecting} decides whether the received speech signal is watermarked by estimating the embedded pattern in two steps (Fig.~\ref{fig:overviewwm}):

\noindent\underline{\ding{175} Watermark extraction.}  
Given a potentially watermarked signal \( x_w \), the detection function \( \mathcal{D}(\cdot) \) outputs an estimated bit sequence
\( w' = \mathcal{D}(x_w) = \{w'_i\}_{i=1}^n \), where \( w'_i \in \{-1, 1\} \) denotes the \( i \)-th predicted bit.

\noindent\underline{\ding{176} Matching score.}
The predicted sequence \( w' \) is evaluated against the watermark \( w \) using the bitwise matching score
$\Theta(w, w') = \frac{1}{n} \sum_{i=1}^{n} \mathbb{I}(w_i = w'_i)$,
where \( \mathbb{I}(\cdot) \) denotes the indicator function, returning 1 if the condition is true and 0 otherwise.

\noindent\textbf{Watermarking decision.}
The detector compares the bitwise matching accuracy with a threshold \(\kappa\) and outputs
\(y \in \{0,1\}\) as \(y = 1\, \text{ if }\, \Theta(w, w') > \kappa,\, \text{else }\, 0\),
where \(y=1\) denotes watermark presence and \(y=0\) absence.
The threshold \(\kappa\) controls the robustness–reliability trade-off~\cite{roman2024proactive,pavlovic2022robust,cvejic2004spread}.

\subsection{Threat Model}\label{sec:threat_model}

\noindent\textbf{Attack Scenario.} 
We consider a scenario in which an adversary attempts to evade detection on watermarked speech to enable downstream misuse, such as unauthorized voice cloning~\cite{liu2023detecting, zongaudiomarknet} or illegal redistribution. The watermark is imperceptible, meaning the attacker is unaware of whether a given audio sample is watermarked, and therefore attempts to disrupt any potential watermark before misuse.

\noindent\textbf{Attack Goal.}
To enable downstream misuse of the audio, the adversary is given a watermarked audio $x_{w_t}$ carrying an unknown target watermark message $w_t$, and produces a perturbed audio $x'$ such that (i) $x'$ evades watermark detection, and (ii) $x'$ remains perceptually close to $x_{w_t}$.

\noindent\textbf{Adversary Capabilities.} 
We adopt a strict zero-query black-box model in which the adversary is constrained along three dimensions: 
(i) No watermarking system knowledge: the adversary has no prior knowledge of the target watermarking system, including its embedding and detection mechanisms. (ii) No query access: the adversary cannot interact with the target watermarking system and receives no feedback during the attack process. Specifically, the target watermarking system is accessed only beforehand to obtain the watermarked audio $x_{w_t}$ as the attack input; once $x_{w_t}$ is obtained, zero queries are made to the target detector throughout the attack. (iii) No access to clean or watermarked dataset: the adversary only has access to the single target watermarked speech $x_{w_t}$ to be attacked, and no access to any additional clean speech $x$ or watermarked data from the target speaker. As a result, the attacker cannot exploit speaker-specific features such as timbre. Instead, the adversary relies on a publicly available surrogate watermarking model, which is not assumed to match the target system and is used only to learn transferable perturbations $\delta$. Therefore, our zero-query constraint applies specifically to the target detector during the attack, while allowing access to the surrogate model.

\noindent\textbf{Defender's Knowledge and Capabilities.} The defender is the party that embeds watermarks into speech to support provenance verification and misuse detection. The defender has full control over the watermarking system and can deploy any watermarking model unknown to the adversary. Following the principle of adaptive threat modeling~\cite{yusmack,li2023inaudible,wang2024fraudwhistler}, we further assume that the defender is aware of the adversary's algorithm, including the optimization strategy used to generate $\delta$. The defender leverages this knowledge to develop adaptive countermeasures, such as recovering watermark evidence on suspicious speech samples.

\begin{table}[t]
\centering
\caption{Adversary's knowledge and evasion scope.}
\label{tab:attack_knowledge}

\setlength{\tabcolsep}{2pt}

\begin{threeparttable} 
\resizebox{0.98\columnwidth}{!}{
\begin{tabular}{@{} l cc cc cc @{}}
\toprule
\multirow{2}{*}{\textbf{Method}} & \multicolumn{2}{c}{\textbf{System Access}} & \multicolumn{2}{c}{\textbf{Data Dependencies}} & \multicolumn{2}{c}{\textbf{Evasion Scope}} \\
\cmidrule(lr){2-3} \cmidrule(lr){4-5} \cmidrule(l){6-7} 
& \textbf{WM Pipeline} & \textbf{Query} & \textbf{Paired Data} & \textbf{Target Data} & \textbf{Academic} & \textbf{Commercial} \\
\midrule
White-box Noise~\cite{liu2024audiomarkbench}  & \cmark & \cmark & \xmark & \xmark & \cmark & \xmark \\
Black-box Noise~\cite{liu2024audiomarkbench}  & \xmark & \cmark & \xmark & \xmark & \cmark & \xmark \\
Overwrite (query)~\cite{yao2026yours}         & \xmark & \cmark & \xmark & \xmark & \cmark & \xmark \\
Overwrite$_{NQ}$~\cite{yao2026yours}     & \xmark & \xmark & \xmark & \xmark & \cmark & \xmark \\
SE Attack~\cite{lopez2024speech}              & \xmark & \xmark & \cmark & \cmark & \cmark & \xmark \\
HarmonicAttack~\cite{li2025harmonicattack}    & \xmark & \xmark & \cmark & \cmark & \cmark & \xmark \\
DiffErase~\cite{yaoblack}                     & \xmark & \xmark & \xmark & $\triangle$ & \cmark & \xmark \\

Encodec~\cite{defossez2022encodec}                   & \xmark & \xmark & \xmark & $\triangle$ & \cmark & \xmark \\
\midrule
\rowcolor{gray!15} 
\textbf{ScrambleMark}         & \textbf{\xmark} & \textbf{\xmark} & \textbf{\xmark} & \textbf{\xmark} & \textbf{\cmark} & \textbf{\cmark} \\
\bottomrule
\end{tabular}%
}
\vspace{2pt}
\begin{tablenotes}[flushleft]
\footnotesize
\item \parbox{0.59\linewidth}{
(i) \textbf{WM Pipeline}: knowledge of the watermark embedding and detection pipeline~\cite{liu2024audiomarkbench};
(ii) \textbf{Query}: query access to the watermark detector~\cite{chen2020hopskipjumpattack,andriushchenko2020square}, and Overwrite$_{NQ}$ denotes the non-query variants of \cite{yao2026yours}; 
(iii) \textbf{Paired Data}: paired clean and watermarked speech~\cite{lopez2024speech,li2025harmonicattack}; 
(iv) \textbf{Target Data}: clean target-speaker speech samples. DiffErase~\cite{yaoblack} and recent works~\cite{defossez2022encodec, o2025deep, wen2025sok,yaoblack} do not clarify whether the pretrained generative model has been exposed to the clean target speaker's speech. We denote this assumption as $\triangle$ to reflect the ambiguity, and revisit it empirically in Section~\ref{sec:evaluation};
(v) \textbf{Evasion Scope}: evaluated on \textbf{Academic} schemes and/or \textbf{Commercial} platforms.
}
\end{tablenotes}
\end{threeparttable}
\end{table}

\begin{figure*}[t]
    \centering \includegraphics[width=1.0\linewidth]{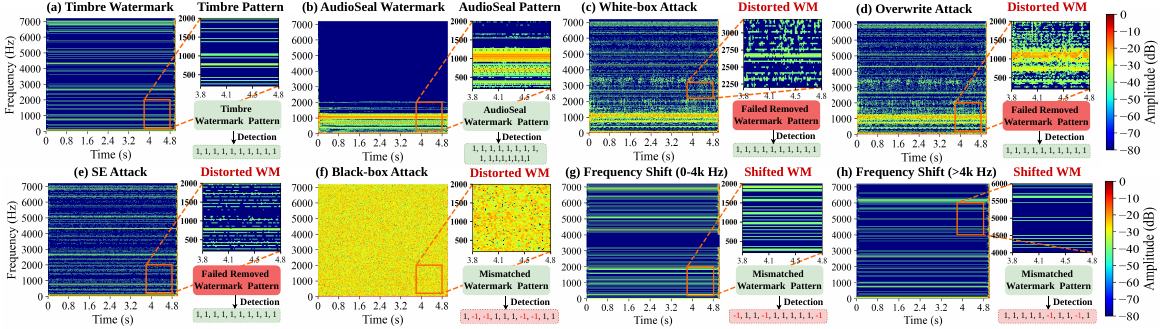}
\caption{Comparison of watermark patterns under different attack strategies. The Timbre watermark~\cite{liu2023detecting} is used as the target, while AudioSeal~\cite{roman2024proactive} serves as a surrogate watermarking model to which noise-based~\cite{liu2024audiomarkbench}, overwrite~\cite{yao2026yours}, and SE~\cite{lopez2024speech} attacks are applied. (a) Timbre watermark and (b) AudioSeal pattern under the same all-ones message; (c)--(f) patterns after existing attacks; (g)--(h) Simple scrambling: (g) 0--4 kHz shifted down by 753 Hz; (h) above 4 kHz shifted down by 753 Hz.}
\label{fig:compareattack}
\end{figure*}

\subsection{Limitation of Existing Attacks}\label{relatedwork}
Existing watermark evasion attacks can be broadly categorized into three types: noise-based attacks, overwriting attacks, and generative removal attacks. (i) Noise-based attacks~\cite{liu2024audiomarkbench, chen2020hopskipjumpattack, andriushchenko2020square} perturb the watermarked audio to suppress detection, either through white-box optimization with full knowledge of the embedding and detection pipeline, or through black-box queries to the detector. (ii) Overwriting attack~\cite{yao2026yours} instead embeds surrogate watermarks into the target audio to overwrite the original, relying on either detector feedback or brute-force surrogate ensembles. (iii) Generative removal attacks~\cite{lopez2024speech, li2025harmonicattack, o2025deep, wen2025sok,yaoblack} leverage generative models to reconstruct unwatermarked audio, and split into supervised approaches that require paired clean--watermarked data~\cite{lopez2024speech, li2025harmonicattack} and unsupervised approaches that rely on pretrained generative priors~\cite{yaoblack}. Table~\ref{tab:attack_knowledge} summarizes the knowledge each attack assumes about the watermarking system. 

However, existing attacks make assumptions that may not hold under realistic threat models. Noise-based and overwriting attacks typically require white-box access or detector queries, while supervised generative attacks rely on paired clean--watermarked data that is impractical to obtain at scale. Unsupervised generative attacks such as DiffErase~\cite{yaoblack} avoid these requirements, but their reliance on pretrained generative models (e.g., BigVGAN vocoder~\cite{lee2022bigvgan}) introduces an implicit dependence on the pretraining distribution: On audio from unseen target speakers, the generative prior fails to generalize and produces audible distortions, as we demonstrate in Section~\ref{sec:realworld}.

\noindent\textbf{Limited Transferability.}
To understand why existing attacks transfer poorly across watermarking schemes, we compare their perturbation spectra in Fig.~\hyperref[fig:compareattack]{\ref{fig:compareattack}}, using AudioSeal as the surrogate and Timbre as the target. Three observations emerge from the comparison: (i) The white-box attack (Fig.~\hyperref[fig:compareattack]{\ref*{fig:compareattack}(c)}) injects localized low-frequency perturbations tailored to AudioSeal's watermark pattern (Fig.~\hyperref[fig:compareattack]{\ref*{fig:compareattack}(b)}). Because Timbre exhibits a different watermark structure (Fig.~\hyperref[fig:compareattack]{\ref*{fig:compareattack}(a)}), the perturbation only partially suppresses detection; (ii) The overwrite attack (Fig.~\hyperref[fig:compareattack]{\ref*{fig:compareattack}(d)}) concentrates its signal below 1200~Hz, leaving most of Timbre's watermark structure unaffected; (iii) The SE attack (Fig.~\hyperref[fig:compareattack]{\ref*{fig:compareattack}(e)}) produces sparse, dot-like perturbations that affect only isolated portions of Timbre's stripe pattern (Fig.~\hyperref[fig:compareattack]{\ref*{fig:compareattack}(a)}), leaving its overall structure largely intact. Despite their different mechanisms, all three attacks target localized watermark artifacts rather than the broader structure used for verification, thus transfer poorly to unseen watermarking systems.

\noindent\textbf{Perceptible Distortion.} A simple way to improve transferability is to enlarge the perturbation budget to further disrupt the watermark pattern. With a high budget, the black-box attack in Fig.~\hyperref[fig:compareattack]{\ref*{fig:compareattack}(f)} bypasses detection by flipping 40\% of the watermark bits, but at a clear cost: its perturbation energy exceeds the white-box attack by 20--30~dB, producing a dense, noisy spectrogram and rendering the speech unusable. In practice, the resulting audio is no longer usable (e.g., voice cloning) due to severe perceptual degradation.

\subsection{Design Intuition}\label{sec:design}

A natural evasion strategy is to locate and remove watermark artifacts, but this is rarely feasible: watermarks are designed to be imperceptible, dispersed across time~\cite{roman2024proactive} or confined to narrow frequency bands~\cite{liu2023detecting}, and masked by the host speech itself. Without clean reference speech, their precise locations cannot be reliably inferred.

The difficulty of locating watermark artifacts raises a fundamental question: \emph{is removal even necessary for evasion?} The answer lies in how watermarks are constructed. Modern watermarking systems~\cite{liu2023detecting,roman2024proactive,zongaudiomarknet} encode messages as structured patterns designed to survive common distortions such as compression and noise. To achieve both robustness and capacity, watermark information is typically distributed across the time-frequency representation of the signal rather than embedded at a single location. The form of this distributed watermark pattern commonly differs across methods; e.g., Timbre and AudioSeal produce markedly different spectro-temporal layouts (Fig.~\hyperref[fig:compareattack]{\ref*{fig:compareattack}(a)–(b)}), and detection likely depends on matching these watermark-dependent patterns. This observation suggests a different strategy for watermark evasion: rather than removing watermark artifacts, an adversary can perturb the time-frequency structure of speech to disrupt this consistency and cause detection to fail. Since this strategy operates directly on the watermarked signal without requiring access to or modeling of the watermark model, it is particularly suitable for real-world black-box settings where the target system is inaccessible.

\noindent\textbf{Scrambling Observation.}
To assess whether such direct perturbation is feasible, we probe with a simple and interpretable scrambling operation: band-wise frequency shifting, which warps spectral patterns within fixed frequency bands. We find that frequency shifts corrupt 30\% of watermark bits in low-frequency bands and 20\% in high-frequency bands (Fig.~\ref{fig:compareattack}(g)--(h)). Although the watermark artifact is not explicitly erased, detection accuracy drops substantially, establishing structural disruption as a viable attack strategy. Next, we build on this insight to develop a new framework for structural watermark evasion.

\subsection{The ScrambleMark Framework}
We propose ScrambleMark, a zero-query, watermark-agnostic evasion attack that defeats detection by disrupting structural consistency. Realizing this idea in a strict black-box setting raises two challenges. \textit{(i) Full-spectrum coverage under perceptual constraints.} A watermark may reside anywhere in the spectrum, as watermark footprints vary substantially across schemes~\cite{liu2023detecting,roman2024proactive}. Without knowing the watermark's location, an effective attack must provide full-spectrum coverage, remaining capable of disrupting watermark structures regardless of where they are embedded. At the same time, indiscriminate perturbations can distort perceptually important speech characteristics such as pitch and timbre. Effective evasion therefore requires achieving two competing objectives: full-spectrum coverage and perceptual quality preservation. \textit{(ii) Recovery resistance.} Deterministic signal-processing transforms such as frequency shifting can effectively disrupt watermark structure, but their fixed transformation rules leave predictable perturbation patterns. An adaptive defender may estimate these patterns and reverse the attack to restore watermark evidence. Robust evasion therefore requires achieving resistance to recovery.

\begin{figure}[t]
    \centering
    \includegraphics[width=0.47\textwidth]{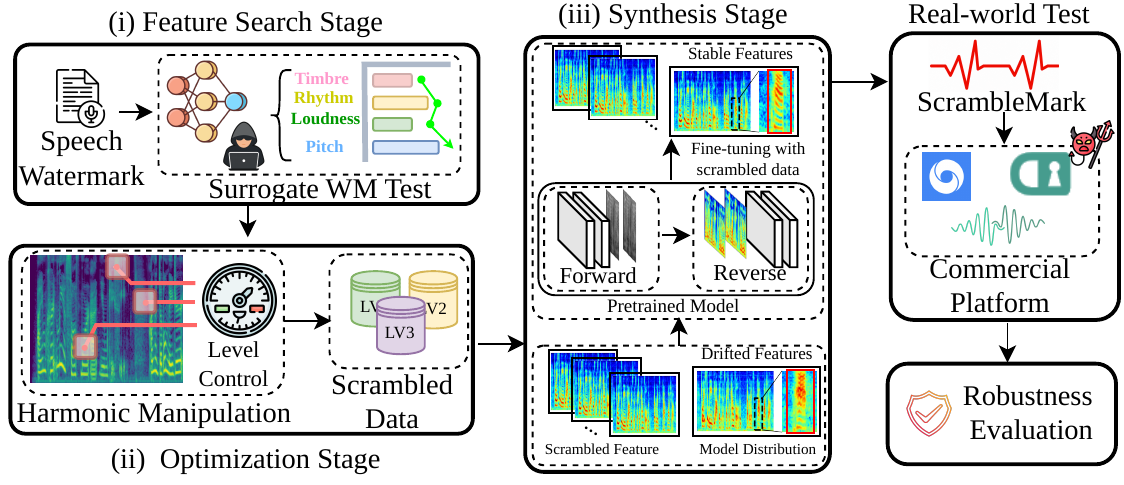}
    \caption{Overview of ScrambleMark.}
    \vspace{-0.0cm}
    \label{fig:workflow}
\end{figure}

ScrambleMark addresses these challenges through a three-stage pipeline (Fig.~\ref{fig:workflow}). (i) The \emph{search} stage addresses the first challenge by systematically exploring the acoustic feature space to identify a scrambling primitive that balances watermark disruption and perceptual quality (Section~\ref{sec:strategy}). (ii) The \emph{optimization} stage optimizes a deterministic scrambler against a surrogate watermark to apply this primitive, validating its transferability across unseen schemes and generating the training data for generative scrambling (Section~\ref{sec:deterministic}). (iii) The \emph{synthesis} stage addresses the second challenge by lifting these deterministic rules into a learned randomized distribution, producing recovery-resistant scrambling while retaining evasion effectiveness (Section~\ref{sec:scramblemark}). We then evaluate ScrambleMark: Section~\ref{sec:evaluation} measures evasion effectiveness on commercial platforms and reports human perceptual ratings, and Section~\ref{sec:robustness} tests its resilience to adaptive recovery.

\section{Searching for a Feasible Scrambling Space}\label{sec:strategy}
We first formulate watermark evasion and systematically evaluate perceptual feature perturbations. The resulting analysis identifies harmonic disruption as the key evasion mechanism, which motivates the design of ScrambleMark.


\subsection{Problem Formulation}\label{subsec:problem}

Given a watermarked audio $x_{w_t}$ carrying watermark $w_t$, the attacker seeks a perturbation $\delta$ that evades detection while remaining perceptually close to $x_{w_t}$:
\begin{equation}
\min_{\delta}\; \mathcal{L}_{\text{perc}}(x_{w_t} + \delta,\, x_{w_t}) \quad \text{s.t.} \quad y_t = 0,
\label{eq:generalformulation}
\end{equation}
where $\delta$ is the perturbation, $\mathcal{L}_{\text{perc}}$ measures perceptual distance from $x_{w_t}$, and $y_t = 0$ denotes that the target detector fails to recover the watermark from $x_{w_t} + \delta$. Motivated by the structural nature of modern speech watermarks (Section~\ref{sec:design}), we focus on perturbations that can effectively disrupt the speech structure. However, the space of possible scrambling operations remains vast, and different perturbations may affect watermark detection and perceptual quality in fundamentally different ways. We next seek effective scrambling directions that balance evasion and fidelity.

\subsection{Perceptual Features as the Scrambling Space}\label{sec:feature_effectiveness}We explore scrambling directions in a perceptual feature space~\cite{li2024safeear,liu2023unifyspeech}, characterized by \textit{pitch}, \textit{rhythm}, \textit{timbre}, and \textit{loudness}. This choice is motivated by two considerations. (i) The perceptual feature space is compact and interpretable, allowing us to pinpoint the perceptual factors responsible for watermark evasion and to explicitly control the evasion–quality trade-off by tuning feature-level scrambling strength. (ii) Compared to localized noise injection, feature-level modifications naturally tend to propagate across broad time-frequency regions. Such full-spectrum coverage directly addresses the challenge that watermark may reside anywhere in the spectrum, enabling comprehensive disruption of watermark structures while preserving the fidelity.

\subsubsection{Feature-level Perturbation Dataset Setup}

To systematically probe watermark behavior under feature-level perturbations, we construct an evaluation dataset along two functional axes: (i) perturbation type, covering a set of perceptually meaningful features whose effects jointly span complementary regions of the spectrogram, and (ii) perturbation strength, sampled across a wide range of intensities to characterize the evasion--quality trade-off. The corresponding features are introduced below.

\noindent\textbf{Pitch} determines harmonic structure~\cite{moore2012psychology,mcdermott2008music}. We consider (i) \emph{Pitch shifting}, which changes pitch by scaling harmonic structure; (ii) \emph{Frequency shifting}, which shifts spectrogram energy along the frequency axis rather than scaling it.

\noindent\textbf{Rhythm} governs temporal structure~\cite{large1999dynamics,patel2014evolutionary}. We consider (i) \emph{Time compression/expansion}, which alters tempo; (ii) \emph{Time cropping}, which removes a signal segment.

\noindent\textbf{Timbre} characterizes the spectral envelope of a sound~\cite{mcadams1995timbre}. We consider (i) \emph{Quantization}, which introduces quantization noise; (ii) \emph{Median filtering}, which smooths spectral details.

\noindent\textbf{Loudness} controls overall signal intensity~\cite{zwicker2007psychoacoustics,iturbs1770}. We consider \emph{amplitude scaling} to vary signal energy.

\noindent\textbf{Baselines.} We compare feature-level scrambling against three baselines: (i) additive Gaussian noise; (ii) ambient sound injection~\cite{deng2022fencesitter} (e.g., car); and (iii) speech enhancement~\cite{ephraim1984speech} and denoising~\cite{berouti1979enhancement}, applied as a preprocessing step that aims to suppress watermark-related components.

\begin{table}[t]
\caption{Parameters for each perturbation method.}
\centering 
\renewcommand{\arraystretch}{0.90}
\resizebox{0.43\textwidth}{!}{%
\begin{tabular}{c|cccccc} 
\hline
\multicolumn{7}{c}{}\\[-1.0em]

\multirow{3}{*}{\textbf{Method}} & & \multicolumn{5}{c}{\textbf{Perturbation Strengths}} \\ \cline{3-7}
\\[-1.0em]
 & & \textbf{Level 1} & \textbf{Level 2} & \textbf{Level 3} & \textbf{Level 4} & \textbf{Level 5} \\
\multicolumn{7}{c}{}\\[-1.0em]
\hline
\multicolumn{7}{c}{}\\[-1.0em]

\textbf{Median Filtering (samples)} & &5  & 15  & 25 & 35 & 45 \\
\rowcolor{gray!20} \textbf{Quantization (bps)}& & 9 & 8 & 7 & 5 & 3 \\

\textbf{Gaussian Noise (dB)}& & 25 & 20 & 15 & 10 & -10 \\


\rowcolor{gray!20} \textbf{Downsampling (HZ)}& & 16000 & 14000 & 6000 & 4000 & 2500 \\
\textbf{Background Noise (dB)}& & 25 & 20 & 15 & 10 & -10 \\

\rowcolor{gray!20}\textbf{Amplitude Scaling (\%)}& & 60 & 40 & 10 & 5 & 0.01 \\

\textbf{Pitch Shift Up (Cents)}&  & 15 & 35 & 200 & 300 & 400 \\

\rowcolor{gray!20}\textbf{Pitch Shift Down (Cents)}& & 15 & 35 & 200 & 300 & 400 \\

\textbf{Frequency Shift (HZ)}& & 10 & 15 & 20 & 25 & 30 \\





\rowcolor{gray!20}\textbf{Time Crop(\%)}& & 20 & 40 & 80 & 95 & 99 \\

\textbf{Time Compression}& & 0.8  & 0.6  & 0.4  & 0.2  & 0.1  \\

\rowcolor{gray!20}\textbf{Time Expansion}& & 1.25  & 1.67  & 2.5  & 5  & 10
\\

\multicolumn{7}{c}{}\\ [-1.0em]
\hline
\end{tabular}
}
\label{tab:distortionparameter}
\end{table}

\subsubsection{Experimental Setup}
(i) Perturbation Configuration: To cover a broad range of perturbation strengths, we configure each method at five levels (Level~1--5), with parameters summarized in Table~\ref{tab:distortionparameter}. Existing speech enhancement and denoising methods are applied as fixed preprocessing steps and therefore have no level settings. (ii) Dataset: We randomly sample 500 speech segments each from Valentini~\cite{valentini2017noisy} and VoxCeleb1~\cite{nagrani2017voxceleb} (1{,}000 clips total), and apply every method at all five levels to construct the evaluation set.

\noindent\textbf{Target Watermarking Model.} For the experiments in this section, we adopt Timbre~\cite{liu2023detecting}, a robust watermark designed to stay detectable under common audio processing. We use it as a demanding target because it spreads its watermark across nearly the entire spectrum, so evading it requires a perturbation that spans the full spectrum, matching the requirement we identified for effective evasion.


\noindent\textbf{Evaluation Metric.} To identify which feature manipulation achieves strong evasion with minimal speech degradation, we define the \emph{Evasion--Quality Trade-off Ratio} (EQTR), which measures the drop in watermark matching score per unit of perceptual quality loss~\cite{dong2018metric,duanparrot}:
$\mathrm{EQTR} = \Delta\Theta\,(2 - \mathrm{STOI})^{-1}$,
where $\Delta\Theta$ is the drop in watermark matching score, computed as the score on the original watermarked speech minus the score on the perturbed speech (both normalized to $[0,1]$). Short-Time Objective Intelligibility (STOI)~\cite{taal2011algorithm} measures speech intelligibility, with higher values indicating better perceptual quality. The denominator $(2-\mathrm{STOI})$ normalizes the ratio: since practical STOI values typically lie in $[0,1]$, $(2-\mathrm{STOI})$ remains bounded away from zero, and confines EQTR to $[0,1]$. A larger EQTR indicates stronger evasion at lower perceptual cost.

\subsubsection{Results Analysis}\label{sec:Mechanisms}
\begin{figure}[t]
    \centering
    \includegraphics[
        width=0.98\columnwidth,
        height=99.45566pt
    ]{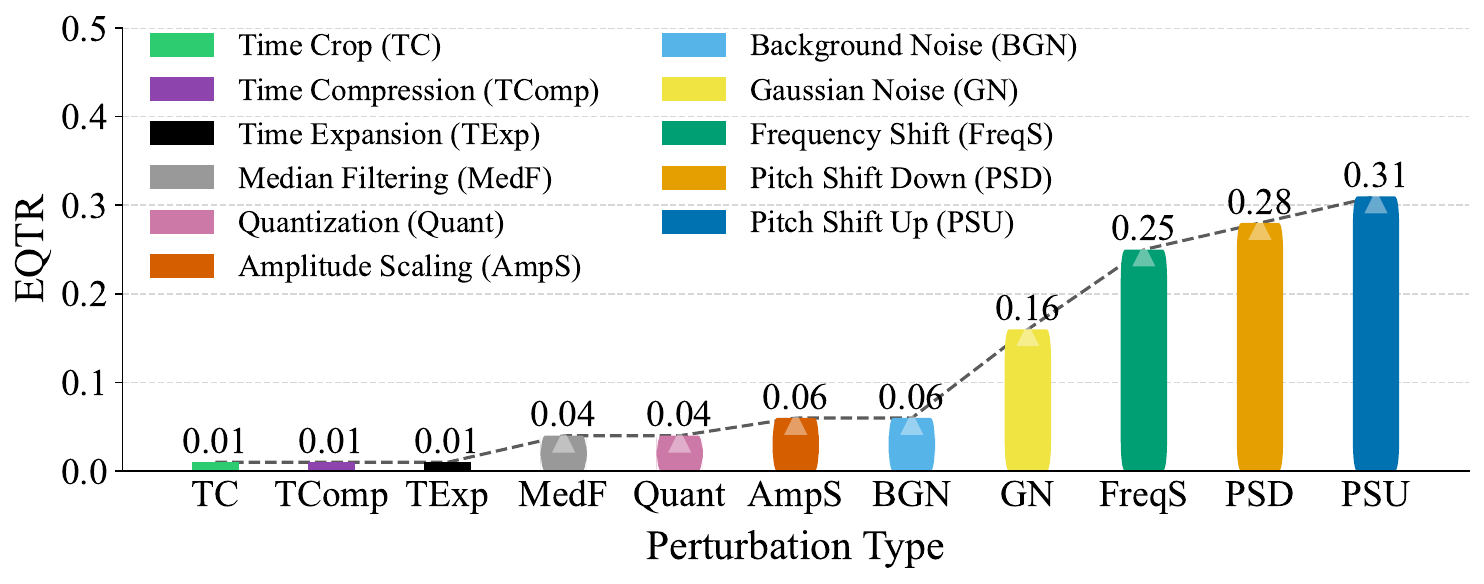}
    \caption{EQTR across feature-level perturbations.}
    \label{fig:EQTR}
\end{figure}

Fig.~\ref{fig:EQTR} reveals a clear advantage of harmonic-axis perturbations. Pitch shifting up \& down, and frequency shifting achieve the highest EQTR scores (0.31, 0.28, and 0.25), substantially outperforming Gaussian noise (0.16), rhythm-based perturbations, and denoising methods (both near 0.01). The common property shared by these strongest attacks is harmonic disruption. Pitch shifting rescales every harmonic component by a shared ratio; frequency shifting moves them by a shared offset. Both transform the harmonic structure as a coherent whole. The weaker methods touch other axes: timbre-based operations (e.g., quantization) perturb local spectral texture, rhythm-based methods alter temporal structure, and speech enhancement/denoising removes noise-like artifacts, all leaving the harmonic structure intact.

\noindent\textbf{Core Mechanism of Structural Evasion.}
\textit{Watermark evasion is primarily driven by the targeted disruption of harmonic structures.}
This finding motivates a scrambling strategy based on controlled frequency-axis deformation. Instead of injecting generic noise or altering temporal structures, the scrambling mechanism should stretch, compress, or warp harmonic components to break watermark-specific structural consistency while preserving perceptual quality.

\subsection{From Harmonic Scrambling to ScrambleMark}
While the above strategy is promising, turning it into a practical generative ScrambleMark framework faces an immediate difficulty: training the generator requires a dataset of harmonically scrambled speech, and no public dataset is tailored to this purpose. To close this gap, we need a scalable data generation pipeline that produces scrambled samples achieving both high evasion and good perceptual quality. Accordingly, we structure our approach as follows. \textit{(i) A Deterministic Data Engine.} We first build a deterministic harmonic scrambler that serves as a data engine, supplying the training datasets the generator needs. We study and optimize a deterministic scrambling operation that generalizes across multiple watermarking models. \textit{(ii) Generative ScrambleMark Training.} We then use this optimized scrambler to construct the training dataset for a generative ScrambleMark model. The resulting model preserves the evasion effectiveness, while introducing randomness that improves robustness against adaptive recovery and defenses.

\section{Building the Deterministic Data Engine}\label{sec:deterministic}
This section develops a deterministic scrambler for generating ScrambleMark training data. We first construct a band-wise harmonic scrambler, then optimize it against a surrogate watermark detector and evaluate its effectiveness across diverse watermarking models.

\noindent\textbf{Band-wise Harmonic Scrambling Formulation.}\label{subsec:formulation}
Since watermark signatures and perceptual sensitivity are distributed unevenly across the frequency spectrum, global transformations (e.g., uniform pitch shifting) inevitably force a suboptimal trade-off between evasion and audio fidelity. To address this, our deterministic scrambler partitions the entire frequency axis into multiple bands, applying independent frequency shifts to each. Specifically, given a watermarked signal $x_{w_t}$, the scrambler leverages a band-wise signal-processing operator $\mathcal{H}$ to generate the scrambled output:
\[
x_{w_t}' = \mathcal{H}\!\left(x_{w_t} \mid (\boldsymbol{\Omega}, \mathbf{S})\right),
\]
where the operator $\mathcal{H}$ is implemented using a modified open-source framework~\cite{pythonstretch} that extends conventional global spectral transformations to flexible band-wise manipulations. Formally, the deterministic scrambler is parameterized by two components: a frequency partition $\boldsymbol{\Omega}$ and a shift vector $\mathbf{S}$. The partition splits the frequency axis into $K$ contiguous bands $\boldsymbol{\Omega}=\{\Omega_k\}_{k=1}^K$, whose widths $|\Omega_k|$ need not be equal. Each band $\Omega_k$ then receives a signed shift $s_k \in \mathbb{R}$ in cents, positive for upward and negative for downward, collected into $\mathbf{S}=\{s_k\}_{k=1}^K$. Together, $(\boldsymbol{\Omega}, \mathbf{S})$ determine how the spectrum is scrambled in frequency-axis.

\begin{figure}[t]
    \centering
    \includegraphics[width=0.98\linewidth]{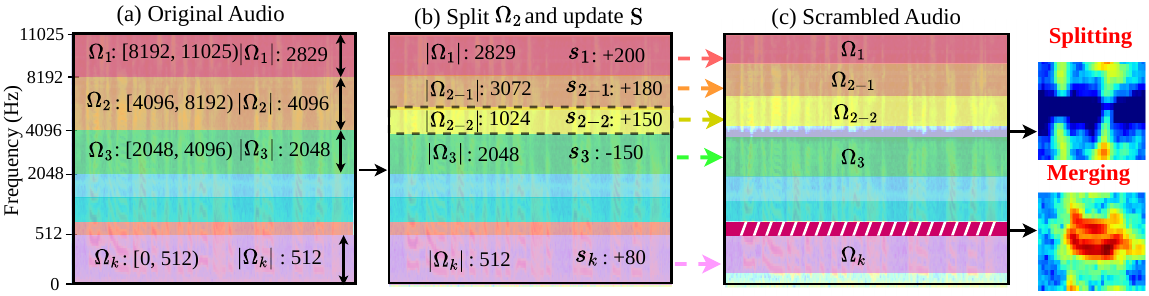}
    \caption{Harmonic partition-shift operation.}
    \label{fig:partition}
\end{figure}



\subsection{Surrogate-Guided Optimization}
In practice, the target detector is unknown, so we cannot optimize the scrambler against it directly. We instead optimize against a surrogate: we embed a surrogate watermark $w_s$ into the audio via $x_s = \mathcal{E}(x_{w_t}, w_s)$, then optimize the scrambling parameters to suppress $w_s$. Because embedded watermarks are imperceptible by design, this surrogate embedding leaves the audio essentially unchanged. We optimize the scrambling parameters $(\boldsymbol{\Omega}, \mathbf{S})$ by replacing the hard detection constraint in \eqref{eq:generalformulation} with a soft surrogate penalty:
\begin{equation}
\begin{aligned}
\arg\min_{\boldsymbol{\Omega},\mathbf{S}} \quad & \Theta\!\left(\mathcal{D}(x_s'),\, w_s\right) + \lambda\,\mathcal{L}_{\text{Mel}}(x_s',\, x_{w_t}) \\
\text{s.t.} \quad & x_s' = \mathcal{H}\!\left(x_s \mid (\boldsymbol{\Omega},\mathbf{S})\right),
\end{aligned}
\label{eq:problemformulation1}
\end{equation}
where the first term, $\Theta(\cdot,\cdot)$, drives the surrogate detector to fail on the scrambled audio, and the second realizes $\mathcal{L}_{\text{perc}}$ as a Mel-spectrogram distance $\mathcal{L}_{\text{Mel}}(\cdot,\cdot)$ that keeps the audio perceptually close to the original \cite{wang2024vibspeech,zhang2025safespeech}. The weight $\lambda$ balances evasion against fidelity.

\noindent\textbf{Alternating Partition and Shift Optimization.} The optimization problem in \eqref{eq:problemformulation1} jointly searches for the frequency partition $\boldsymbol{\Omega}$ and the corresponding shift vector $\mathbf{S}$. However, this optimization is challenging because $\boldsymbol{\Omega}$ and $\mathbf{S}$ are coupled: modifying $\boldsymbol{\Omega}$ changes the number and boundaries of frequency bands, thereby altering the dimensionality of $\mathbf{S}$. Our idea is to break this coupling by alternating (Fig.~\ref{fig:partition}): we fix the partition $\boldsymbol{\Omega}$ and optimize the shifts $\mathbf{S}$ under it, then split the wider bands into narrower ones and repeat. Each split gives the next round of shift optimization a finer partition to work on. As the bands grow narrower, each shift acts on a smaller slice of the spectrum, letting the scrambler balance evasion and quality more finely.


Algorithm~\ref{alg:BA} realizes this idea. We initialize the partition and shifts (Line 1) and the objective function $\mathcal{J}(\cdot,\cdot)$ (Line 2). Each iteration first refines the partition, splitting bands with probability proportional to their width so wider bands divide more frequently (Lines 3--7); each sub-band inherits the shift of the band it split from, keeping the scrambling continuous (Line 8). The shift vector is then optimized under the refined partition using the Covariance Matrix Adaptation Evolution Strategy (CMA-ES)~\cite{hansen2001cmaes} (Line 9). The resulting configuration is retained only if it reduces the objective (Line 11). The best configuration produces the scrambled audio (Line 12).

\begin{algorithm}[t]
\caption{Partition-Shift Optimization}
\label{alg:BA}
\SetCommentSty{footnotesize}
\SetKwInOut{Input}{Input}
\SetKwInOut{Output}{Output}
\SetKwFunction{Scramble}{Scramble}
\SetKwFunction{Loss}{Loss}
\SetKwFunction{CMAES}{CMA-ES}
\Input{surrogate audio $x_s$; initial partition $\boldsymbol{\Omega}^0$ and shifts $\mathbf{S}^0$; outer iterations $T$; CMA-ES iterations $M$; candidate count $z$}
\Output{scrambled audio $x_s'$}
\BlankLine
$\boldsymbol{\Omega}, \mathbf{S} \gets \boldsymbol{\Omega}^0, \mathbf{S}^0$;\quad $\mathcal{J}^* \gets \infty$\;
$\mathcal{J}\gets \Theta\!\left(\mathcal{D}(x_s'),\, w_s\right) + \lambda\,\mathcal{L}_{\text{Mel}}(x_s',\, x_{w_t})$;
\BlankLine
\For{$t \gets 1$ \KwTo $T$}{
  \tcp{Refine the partition: split a few of the widest bands}
  $\mathcal{K} \gets$ selecting the $z$ widest bands in $\boldsymbol{\Omega}$\;
  \ForEach{$\Omega_k \in \mathcal{K}$}{
  \textbf{with probability} $P_k = |\Omega_k| / \max_j |\Omega_j|$:\\
randomly split $\Omega_k$ into two sub-bands;\\
each sub-band inherits the shift $s_k$ of $\Omega_k$\;
    }
  \tcp{Optimize the shifts under the refined partition}
  $\mathbf{S} \gets \CMAES(\boldsymbol{\Omega}, \mathbf{S}, M)$ \tcp*{minimize $\mathcal{J}$, $\boldsymbol{\Omega}$ fixed}
  $\mathcal{J}\gets\mathcal{J}(\boldsymbol{\Omega}, \mathbf{S})$\\
  \BlankLine
  \lIf{$\mathcal{J} < \mathcal{J}^*$}{
    $\boldsymbol{\Omega}^*, \mathbf{S}^*, \mathcal{J}^* \gets \boldsymbol{\Omega}, \mathbf{S}, \mathcal{J}$
  }
}
\BlankLine
\Return $x_s' \gets \mathcal{H}\!\left(x_s \mid (\boldsymbol{\Omega^*},\mathbf{S^*})\right)$\;
\end{algorithm}
\subsection{Evasion and Quality Evaluation} \label{sec:al-evasionsetup}
We next evaluate whether the optimized scrambler achieves strong cross-model transferability to unseen watermarking models while maintaining perceptual quality.

\begin{figure*}[!t]
    \centering
    \includegraphics[width=1.0\linewidth]{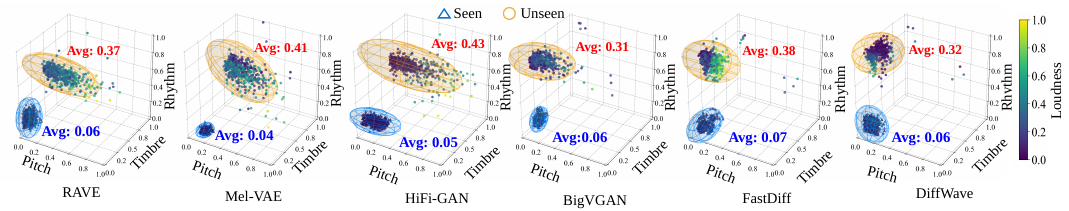}
    \caption{Feature-variation distributions on seen vs.\ unseen data. Over 500 original clips, we measure variations in loudness (Root Mean Square difference~\cite{skovenborg2004evaluation}), pitch ($F_0$ difference~\cite{jadoul2018introducing}), timbre (Mel-Cepstral Distortion~\cite{kubichek1993mel}), and rhythm (Energy-Envelope Dissimilarity~\cite{foote2002audio}), all normalized to $[0,1]$ (lower indicates less variation). Details are provided in Appendix~\ref{feature}.}
    \label{fig:visualizationfourfeaturedistribution}
\end{figure*}

\noindent\textbf{Datasets.} We randomly select 500 test samples from VoxCeleb1~\cite{nagrani2017voxceleb} and VOiCES~\cite{richey2018voices}, and embed a random watermark message into each sample.

\noindent\textbf{Watermarking methods.} We evaluate against three state-of-the-art watermarking methods that span the time and frequency domains: AudioSeal, WavMark, and SilentCipher, each run with its official implementation.

\noindent\textbf{Metrics.} We measure evasion effectiveness by the attack success rate (ASR), the fraction of perturbed samples that evade the target detector, and perceptual quality by STOI.

\noindent\textbf{Implementation.} We use Timbre as the surrogate watermark model for optimization. The partition $\boldsymbol{\Omega}^0$ is initialized as a single band spanning the full spectrum, with initial shift $\mathbf{S}^0 = 80$ cents and trade-off weight $\lambda = 1$. We run $T = 50$ outer iterations and $M = 50$ CMA-ES iterations per round. For partitioning, at most $z = 3$ bands are probabilistically selected for splitting in each round.

\begin{table}[ht]
\centering
\renewcommand{\arraystretch}{0.85}
\caption{Evasion Effectiveness and Quality Results.}
\label{tab:initialeffectiveness}
\resizebox{0.48\textwidth}{!}{
\begin{tabular}{lccccccc}
\specialrule{0.10em}{0pt}{0pt}  
\multirow{2}{*}{\textbf{Metric}} & \multicolumn{2}{c}{\textbf{AudioSeal}} & \multicolumn{2}{c}{\textbf{WavMark}} & \multicolumn{2}{c}{\textbf{SilentCipher}} & \multirow{2}{*}{\textbf{Average}} \\

 & VoxCeleb1 & VOiCES & VoxCeleb1 & VOiCES & VoxCeleb1 & VOiCES \\

\midrule
ASR & 0.98 & 1.00 & 1.00& 0.99 & 0.97 & 0.98 & 0.99\\
STOI & 0.91 & 0.92 & 0.90 & 0.90 & 0.91 & 0.90& 0.91\\
\specialrule{0.10em}{0pt}{0pt}
\end{tabular}
}
\end{table}

\noindent\textbf{Evaluation Results.} Table~\ref{tab:initialeffectiveness} shows that the surrogate-tuned scrambler transfers effectively across watermarking methods, achieving nearly 99\% average ASR while maintaining a STOI of 0.91. These results suggest that the learned scrambling strategy captures weaknesses shared across watermarking schemes rather than overfitting to the surrogate model. The deterministic scrambler therefore serves as an effective data engine, providing the training data needed for generative ScrambleMark, which we develop next.

\section{Building Generative ScrambleMark}\label{sec:scramblemark}
We now lift the deterministic scrambler into a generative model, ScrambleMark, that introduces the randomness needed to resist recovery. This transition raises two challenges: (i) how to efficiently learn from deterministic scrambling behaviors, and (ii) how to make the generative model preserve perceptual quality on previously unseen audio. This section presents our solution and integrates its components into the full ScrambleMark attack.

\subsection{Challenges in Generative Scrambling}\label{subsec:challenges}
We model ScrambleMark as a conditional generator that maps the surrogate-watermarked audio $x_s$ to a scrambled output under a scrambling condition $C$: $x' = \mathbb{G}_\theta(x_s \mid C)$. 

\noindent\textbf{A Strawman Solution.}
One possibility is to let the deterministic scrambler produce an optimal condition $C$ and train the generator to reproduce it. However, the generator produces randomized outputs, so a condition that is optimal for the deterministic scrambler may no longer be optimal once reproduced by the generator. Moreover, learning a single deterministic scrambling configuration risks exposing a predictable perturbation pattern that a defender could track and reverse. Instead, the generator should learn from multiple valid conditions that achieve similar effectiveness. Realizing such a generator introduces two challenges.

\noindent\textbf{The Conditional-Space Explosion.}
The space of possible band-wise operations is enormous. Each band carries its own shift, set independently of the others, so the number of full configurations grows exponentially with the number of bands. The solutions in Table~\ref{tab:initialeffectiveness} already use around 100 bands, each shifted within roughly $\pm 40$ to $\pm 80$ cents. Even with only tens of shift values per band, this gives well over $10^{100}$ distinct conditions, far beyond what any training set could enumerate. Finer partitions only push the count higher.

\noindent\textbf{Limited Generalization to Unseen Inputs.}
Our threat model requires the generator to operate on audio it has never seen. This is where unsupervised generative removal attacks tend to break down~\cite{yaoblack}, and ours is no exception: on unseen inputs, the synthesized audio carries audible distortion. To locate its source, we measure how four acoustic features shift from seen to unseen data across generative adversarial networks (GANs), variational autoencoders (VAEs), and diffusion models~\cite{kulkarni2024generalization,kong2020hifi}. The shift is indiscriminate. As Fig.~\ref{fig:visualizationfourfeaturedistribution} shows, all four features drift on unseen inputs, with average changes rising from below 0.1 to 0.31--0.43, roughly five to eight times higher~\cite{chhaglani2025artifree,jiang2025refvc,wang2026neuralcodecs,wang2023gzs}. For an attack that should touch only the harmonic features, every other deviation, in rhythm or loudness, is wasted perceptual budget. Making the generator change only the target features on unseen audio is the second challenge.

\noindent\textbf{Overview.} Fig.~\ref{fig:framework} summarizes the two-phase ScrambleMark pipeline to solve these challenges. \textit{(i) Scramble-domain dataset generation} tackles the conditional-space explosion by compressing band-wise scrambling operations into a small set of intensity-level conditions. During construction, a recovery-aware objective biases the search toward scrambling configurations that are difficult to invert. \textit{(ii) Generative scrambler training} tackles the generalization challenge by training a diffusion-based generator on the resulting dataset. Pretraining provides general synthesis ability, while dynamic fine-tuning enforces harmonic scrambling and output diversity. Finally, ScrambleMark generates multiple intensity-level candidates and retains the minimum-loss one. We next present the design of each component.

\subsection{Scramble-Domain Dataset Generation}\label{subsec:domain}

\noindent\textbf{Intensity-Level Condition.}
To address the conditional-space explosion, instead of mapping each scrambled sample to its own condition, we map a set of samples of similar shift magnitude, differing in their band partitions, to a single intensity-level condition; the collection of these conditions forms the scramble dataset $\mathcal{C}_{\text{scr}}$. This shrinks the conditioning space from per-sample to intensity-level, making training feasible. Because each level pools samples with different band partitions, the generator varies its band structure across outputs rather than reusing a fixed one, making its perturbations harder for a defender to track and reverse. On top of this design, we further harden each sample against recovery and define how each intensity level is built, as detailed below.

\subsubsection{Recovery-Aware Construction}
To make each sample in $\mathcal{C}_{\text{scr}}$ robust against recovery attempts, we exploit a side effect of the partition--shift operation: shifting adjacent bands often causes their spectral energies to overlap and merge (Fig.~\hyperref[fig:partition]{\ref*{fig:partition}(c)}). Once merged, the original band boundaries are lost, and an inverse shift struggles to separate them cleanly. Merging therefore acts as a natural barrier to recovery, though at the cost of some perceptual quality, since merged harmonic energies cannot be cleanly preserved.

We turn this observation into a recovery penalty during the partition-shift search. A defender does not know the per-band shifts $\mathbf{S}$ and can only attempt recovery with a coarse estimate of the overall shift, so we approximate this by averaging $\mathbf{S}$ into a single scalar $\bar{s} = \tfrac{1}{K}\sum_{k=1}^{K} s_k$ and applying $-\bar{s}$ uniformly across all bands: $\tilde{x}_{wt} = \mathcal{H}(x_s' \mid (\boldsymbol{\Omega}, -\bar{s}))$. Detection on $\tilde{x}_{wt}$ then measures how much of the watermark this coarse inversion can recover. We add it as a penalty to the search objective:
\begin{equation}
\mathcal{J} = \Theta\!\left(\mathcal{D}(x_s'),\, w_s\right) 
+ \lambda \mathcal{L}_{\text{Mel}}(x_s', x_{wt})
+ \beta \, \Theta\!\left(\mathcal{D}(\tilde{x}_{wt}),\, w_s\right),
\label{eq:final_loss}
\end{equation}
where the third term, weighted by $\beta$, penalizes configurations whose inversion still recovers the watermark. Configurations with more band merging resist the inversion and incur a smaller penalty, so the search is biased toward them; larger $\beta$ strengthens this bias at the cost of perceptual quality. We update Line~2 of Algorithm~\ref{alg:BA} with \eqref{eq:final_loss}.

\subsubsection{Dataset Generation}
\begin{figure}[t]
    \centering
    \includegraphics[width=0.96\linewidth]{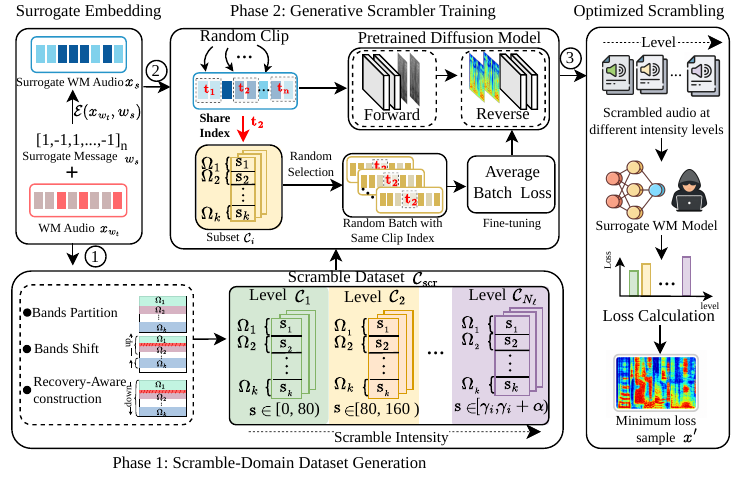}
    \caption{ScrambleMark: dataset generation and training.}
    \label{fig:framework}
\end{figure}

With the recovery-aware objective in place, we run the search under a range of intensity constraints to build $\mathcal{C}_{\text{scr}}$. We index intensity by the magnitude of the shift vector $\mathbf{S} = \{s_k\}_{k=1}^K$: for the $i$-th level, we constrain each shift to $|s_k| \in [\gamma_i,\, \gamma_i + \alpha)$, where $\gamma_i$ is the level's starting magnitude and $\alpha$ a fixed interval width. We run Algorithm~\ref{alg:BA} under each constraint, repeating with different random initializations to diversify the band partitions within a level. As illustrated in Fig.~\ref{fig:framework}, each speech has a subset $\mathcal{C}_i$ of $N_s$ scrambled samples at each intensity level, generated under the same shift range. With $N_{\ell}$ levels, the subsets $\{\mathcal{C}_i\}_{i=1}^{N_{\ell}}$ form the scramble dataset $\mathcal{C}_{\text{scr}}$ with $N_{\ell} \times N_s$ samples per speech.


\subsection{Generative ScrambleMark Training}\label{subsec:training}

We next address the challenge of generalizing harmonic scrambling to unseen audio without introducing unnecessary distortions. Our solution is a pretraining then fine-tuning strategy: pretraining provides general speech synthesis ability, while fine-tuning on $\mathcal{C}_{\text{scr}}$ aligns each scrambling condition with harmonic modifications. Thus, perturbations remain concentrated on harmonic structure for unseen inputs.

\subsubsection{Model and Pretraining}
We seek a generator with non-deterministic outputs, both for perceptual diversity and for the randomized generation that resists recovery. Diffusion models fit on two counts. First, different noise initializations give diverse outputs for the same input, so the scrambling cannot be pinned to a fixed pattern and reversed. Second, the step-by-step denoising operates on fine-grained spectral detail and trains stably on limited data, reconstructing audio at high fidelity. We adopt DiffWave~\cite{kong2021diffwave}, a diffusion vocoder that generates a waveform from a conditioning spectrogram and requires only six inference steps under fast sampling while maintaining high-fidelity speech quality; we drive it with the spectrogram of the input audio so that ScrambleMark maps the surrogate-watermarked audio to its scrambled waveform. To support different intensity levels, a learnable linear layer encodes the input condition, and the resulting embedding is added to the DiffWave spectrogram features as the conditioning input for fine-tuning.

\subsubsection{Dynamic Fine-Tuning}
We now fine-tune the model on $\mathcal{C}_{\text{scr}}$. The generator should avoid converging to a fixed scrambling pattern that can be reversed. To this end, we introduce two complementary sources of randomness, shown in Fig.~\ref{fig:framework}: randomized supervision across scrambling targets and dynamic temporal scrambling across audio segments.

\emph{Randomized supervision} adds randomness at the sample level. For any input $x_s$, the subset $\mathcal{C}_i$ holds many scrambled samples at the same-level intensity, each one a valid scrambled output the generator may learn to produce. Rather than fix the generator to all of them, each mini-batch randomly draws a small set of 16 and uses them as the generation targets for this update, resampling at the next. The generator learns to produce these samples in the standard diffusion manner: for each one, we add Gaussian noise at a randomly drawn step $r$ of a 50-step schedule and train the model to predict that noise from the noisy sample $x_{s,r}'$, the step $r$, and the input $x_s$, so that it learns to turn $x_s$ into a scrambled output. Averaging the loss over the 16 samples and resampling them each update helps the generator avoid locking onto a fixed scramble pattern.

\emph{Dynamic temporal scrambling} applies this random selection within time. The randomized supervision mechanism draws random targets for the input as a whole; here we instead crop a random temporal segment of the input and draw its targets from the same time range of the scrambled samples from $\mathcal{C}_i$. Because each segment is paired with its own random selection, the scrambling varies across the audio, making it harder to track or reverse, and the generator learns segment-level scrambling rather than one static pattern over the full duration.

\noindent\textbf{Fine-tuning Objective.} Given source speech $x_s$ and its scrambled counterpart $x_s' \sim \mathcal{C}_{\text{scr}}$, where $x_{s,r}'$ denotes the noisy version of $x_s'$ at diffusion step $r$, we fine-tune the generator to predict the injected diffusion noise:
\begin{equation}
\theta^{\star}
=
\arg\min_{\theta}\;
\mathbb{E}_{x_s,\; x_s' \sim \mathcal{C}_{\text{scr}},\;\epsilon,\; r}
\left[
\left\|
\epsilon - \mathbb{G}_{\theta}(x_{s,r}',\, r \mid x_s)
\right\|_1
\right].
\label{eq:diffusion_objective}
\end{equation}
Here, each $x_s'$ is a scrambled sample generated by the deterministic scrambler and stored in $\mathcal{C}_{\text{scr}}$, serving as the supervision target during training. The resulting generator $\mathbb{G}_{\theta^{\star}}$ is then used to perform the attack.

\subsection{Attack Formulation}\label{subsec:attack_objective}
We now assemble the components built in this section into the final attack for \eqref{eq:generalformulation}. Since the target detector is inaccessible, we adopt the surrogate-guided formulation \eqref{eq:problemformulation1}. The attack is formulated as finding the optimal scrambling condition $\mathcal{C}_i$ from the candidate set $\mathcal{C}_{\text{scr}}$:
\begin{equation}
\begin{aligned}
\mathcal{C}_i^{\star} = \arg\min_{\mathcal{C}_i \in \mathcal{C}_{\text{scr}}}\; &\Theta\!\left(\mathcal{D}(x'),\, w_s\right) + \lambda\,\mathcal{L}_{\text{Mel}}(x',\, x_{wt}) \\
\text{s.t.}\; &x' = \mathbb{G}_{\theta^{\star}}(x_s \mid \mathcal{C}_i),
\end{aligned}
\label{eq:problemformulation_final}
\end{equation}
where $\mathcal{C}_i$ is a scrambling condition drawn from the candidate set $\mathcal{C}_{\text{scr}}$, $\Theta$ penalizes detection of the surrogate watermark $w_s$ from $x'$, $\mathcal{L}_{\text{Mel}}$ quantifies perceptual degradation, and $\lambda$ balances evasion and quality. Since $\mathcal{C}_{\text{scr}}$ is a finite set, we solve \eqref{eq:problemformulation_final} via exhaustive search over $\mathcal{C}$, generating $x'$ under each candidate $\mathcal{C}_i$ and selecting the minimizer $\mathcal{C}_i^{\star}$.

\section{Evaluation}\label{sec:evaluation}
Our evaluation proceeds in two parts. (i) Academic schemes: we measure the evasion success and audio quality of ScrambleMark across various academic schemes. (ii) Real-world impact: we attack three commercial platforms and conduct a human study on perceptual quality.
\subsection{Experimental Setup}\label{Sec:expsetup}

\noindent\textbf{Target Watermarks.} We evaluate ScrambleMark against seven academic watermarking schemes spanning the time and frequency domains: Timbre~\cite{liu2023detecting}, AudioSeal~\cite{roman2024proactive}, WavMark~\cite{chen2023wavmark}, Collaborative~\cite{juvela2024collaborative}, AudioMarkNet~\cite{zongaudiomarknet}, RobustDNN~\cite{pavlovic2022robust}, and SilentCipher~\cite{singh2024silentcipher}, all run with their official checkpoints. Timbre serves as our surrogate by default; we also attack it as a target (marked $^*$) using AudioSeal as the surrogate. We keep each scheme's default detection threshold $\kappa$; Table~\ref{tab:threshold} in Appendix~\ref{appendix:Academic} reports each scheme's detection performance under this threshold.

\begin{table*}[!t]
\centering
\caption{ScrambleMark compared with existing attacks on academic watermarking schemes.}
\setlength{\tabcolsep}{4pt}
\renewcommand{\arraystretch}{0.85}
\setlength{\aboverulesep}{0.01ex}
\setlength{\belowrulesep}{0.01ex}
\renewcommand{\arraystretch}{0.85}
\resizebox{0.98\textwidth}{!}{
\begin{tabular}{l cc cc cc cc cc cc cc| cc}
\toprule
\multirow{2}{*}{Method}
& \multicolumn{2}{c}{AudioSeal}
& \multicolumn{2}{c}{WavMark}
& \multicolumn{2}{c}{AudioMarkNet}
& \multicolumn{2}{c}{SilentCipher~}
& \multicolumn{2}{c}{Collaborative}
& \multicolumn{2}{c}{RobustDNN}
& \multicolumn{2}{c}{$\text{Timbre}^{*}$}
& \multicolumn{2}{|c}{\textbf{Average}} \\
\cmidrule(lr){2-3}
\cmidrule(lr){4-5}
\cmidrule(lr){6-7}
\cmidrule(lr){8-9}
\cmidrule(lr){10-11}
\cmidrule(lr){12-13}
\cmidrule(lr){14-15}
\cmidrule(lr){16-17}
& ASR($\uparrow$) & STOI($\uparrow$)
& ASR($\uparrow$) & STOI($\uparrow$)
& ASR($\uparrow$) & STOI($\uparrow$)
& ASR($\uparrow$) & STOI($\uparrow$)
& ASR($\uparrow$) & STOI($\uparrow$)
& ASR($\uparrow$) & STOI($\uparrow$)
& ASR($\uparrow$) & STOI($\uparrow$) 
& ASR($\uparrow$) & STOI($\uparrow$)\\
\midrule
White-Box-PGD~\cite{liu2024audiomarkbench}
& 0.00 & 1.00 & 0.00 & 1.00 & 0.15 & 1.00 & 0.00 & 1.00 & 0.02 & 1.00 & 0.00 & 1.00 & 0.00& 1.00 & 0.02  & 1.00 \\
White-Box-I-FGSM~\cite{liu2024audiomarkbench}
& 0.00 & 1.00 & 0.00 & 1.00 & 0.13 & 1.00 & 0.00 & 1.00 & 0.01 & 1.00 & 0.00 & 1.00 & 0.00& 1.00 &  0.02 &  1.00 \\
Black-Box-HSJA~\cite{liu2024audiomarkbench}
& 0.55 & 0.85 & 1.00 & 0.83 & 0.46 & 0.83 & 1.00 & 0.86 & 1.00 & 0.84 & 0.03 & 0.87 & 0.56 & 0.81 & 0.66  &0.84  \\
Black-Box-S-HSJA~\cite{liu2024audiomarkbench}
& 0.82 & 0.84 & 1.00 & 0.85 & 0.58 & 0.81 & 1.00 & 0.84 & 1.00 & 0.84 & 0.03 & 0.86 & 0.20& 0.83 &  0.66 & 0.84  \\
Overwriting Attack~\cite{yao2026yours}
& 0.00 & 0.98 & 0.03 & 0.97 & 0.14 & 0.99 & 0.27 & 0.98 & 0.37 & 0.98 & 0.02 & 0.97 & 0.00 & 0.98 &  0.12 &  0.98 \\
SE Attack~\cite{lopez2024speech}
& 0.63 & 0.98 & 0.18 & 0.99 & 0.05 & 0.99 & 0.26 & 0.99 & 0.21 & 0.99 & 1.00 & 0.98 & 0.00 & 0.98 & 0.33  & 0.99  \\
Harmonic Attack~\cite{li2025harmonicattack}
& 0.57 & 0.96 & 0.10 & 0.99 & 0.00 & 0.97 &0.17  & 0.96 & 0.13  & 0.98 &0.91  & 0.98 &0.00 & 0.97 & 0.27& 0.97\\
DiffErase-L~\cite{yaoblack}
& 1.00 & 0.70 & 1.00 & 0.72 &0.93  & 0.70 & 1.00  & 0.70 & 1.00 & 0.67 & 0.93 & 0.68 & 1.00& 0.73 & 0.98& 0.70\\
DiffErase-M~\cite{yaoblack}
& 1.00 & 0.72 & 1.00 & 0.71 & 0.95 & 0.71 & 1.00 & 0.73 & 1.00 & 0.69 & 0.95 & 0.70 & 1.00& 0.75 & 0.99 & 0.72\\
Encodec~\cite{defossez2022encodec}
& 0.96  & 0.75 & 1.00 &  0.73 & 0.81 &  0.72 & 1.00 &  0.71 & 1.00 &  0.70 & 0.93 &   0.69 & 0.94& 0.72 & 0.95 & 0.72\\
\midrule
\textbf{ScrambleMark}
& 1.00 & 0.91 & 1.00 & 0.90 & 0.98 & 0.88 & 1.00 & 0.92 & 0.96 & 0.90 & 0.86 & 0.93 & 1.00& 0.91 & \textbf{0.97} & \textbf{0.91}\\
\bottomrule
\end{tabular}}
\label{tab:existattackevaluation}
\end{table*}

\subsubsection{Comparison Attacks} \label{comparisonattack}
The baselines in Table~\ref{tab:existattackevaluation} include (i) noise-based attacks~\cite{liu2024audiomarkbench,madry2017towards,kurakin2018adversarial,chen2020hopskipjumpattack,andriushchenko2020square}, using the default 20-dB SNR bound for white-box attacks and a 10,000-query budget for black-box attacks; (ii) overwriting attacks~\cite{yao2026yours}, using 3 embeddings from the same surrogate with different messages; (iii) supervised generative removal~\cite{lopez2024speech,li2025harmonicattack,hu2020dccrn}; and (iv) unsupervised generative removal~\cite{yaoblack,defossez2022encodec}, both using official default settings. To match our threat model, methods requiring watermark knowledge use Timbre as the surrogate watermarking model, switching to AudioSeal when Timbre is the target.

\noindent\textbf{Dataset.} To eliminate training-data ambiguity mentioned in Table~\ref{tab:attack_knowledge} and enable a fair comparison of feature variation on unseen inputs, all generative models, including ScrambleMark, are trained on LJ Speech~\cite{ljspeech17}. We evaluate them on a disjoint multilingual benchmark spanning English~\cite{nagrani2017voxceleb}, Chinese~\cite{zhou2022emotional}, Korean~\cite{choi_2018}, French~\cite{salesky2021multilingual}, and German~\cite{salesky2021multilingual}, without speaker overlap. For each language, we randomly sample 500 speech clips.

\subsubsection{ScrambleMark Configuration}\label{sec:configuration}

\textit{(i) Scramble-domain dataset generation.}
We construct the scramble dataset using the deterministic scrambler. The evasion--quality weight is set to $\lambda = 1$ and the recovery weight to $\beta = 1$. Alternating Partition--Shift optimization runs for $T = 100$ outer iterations with $M = 50$ CMA-ES iterations per round (out-of-range values are clipped to the corresponding bounds during CMA-ES optimization). We initialize $\boldsymbol{\Omega}^0$ as a single band spanning the full spectrum, set the initial shift to $\mathbf{S}^0 = 80$ cents, and allow at most $z = 3$ bands to be selected for splitting in each round. The resulting dataset contains $N_\ell = 10$ intensity levels with $N_s = 100$ samples per level and a shift-range width of $\alpha = 60$ cents.




\noindent\textit{(ii) Generative training.}
We use DiffWave~\cite{kong2021diffwave} with 30 residual layers and 64 channels. It is pretrained on LJ Speech~\cite{ljspeech17} for $10^5$ steps and fine-tuned on the scramble dataset for $5\times10^4$ steps using Adam ($\mathrm{lr}=2\times10^{-4}$). Fine-tuning uses batch size 16 and a 50-step diffusion schedule.
\noindent\textit{(iii) Attack generation.}
At inference time, we enumerate all $N_\ell=10$ intensity levels and generate one scrambled candidate per level. The candidate minimizing the attack objective is selected as the final output. Detailed inference-time computational is reported in Appendix~\ref{appendix:Academic}.

\noindent\textit{(iv) Surrogate watermarking model.} Same as the comparison attacks, we use Timbre as the surrogate watermarking model and embed a random 10-bit message to produce $x_s$. When Timbre is the target, AudioSeal serves as the surrogate.
 
\subsubsection{Evaluation Metrics} We measure attack effectiveness using ASR, defined as the fraction of perturbed watermarked speeches that fall below the target model's detection threshold, and assess speech quality using STOI.

\subsection{Effectiveness on Academic Schemes}\label{sec:TransferabilityEvaluation}

We first evaluate whether ScrambleMark transfers across seven academic watermarking models and at what cost to audio quality. Results can be found in Table~\ref{tab:existattackevaluation}.

\textit{\underline{Observation 1}: ScrambleMark achieves high evasion effectiveness while preserving speech quality.}
Averaged across all seven watermarking models, ScrambleMark attains an ASR of 0.97 and a STOI of 0.91. Its performance is also consistent across targets, maintaining ASR above 0.86 and STOI above 0.88 on every watermarking model. In contrast, existing attacks achieve at most one objective. White-box attacks, Overwriting, SE, and HarmonicAttack preserve audio quality but attain average ASRs of at most 0.33. Conversely, attacks that evade watermark detection often do so at the expense of perceptual quality. Generative removal methods such as Encodec and DiffErase achieve high ASR but reduce STOI to 0.70--0.75. Together, these results show that ScrambleMark is the only method that simultaneously achieves strong transferability and high perceptual quality.

\noindent\textbf{Detector Threshold Sensitivity.}
We further examine ScrambleMark under relaxed detector thresholds with higher FPRs. As shown in Table~\ref{tab:thresholdsensitivity}, ScrambleMark maintains high ASR across increasingly relaxed thresholds. Detailed setup and results are provided in Appendix~\ref{thresholdsensitivity}.

\subsubsection{Impact of Surrogate Choice}

We evaluate whether ScrambleMark remains effective under different surrogate watermarking models. To this end, we rotate the surrogate across all academic watermarking models considered in our evaluation. We use the same multilingual benchmark as before, randomly sampling 500 speech clips per language.

\begin{figure}
    \centering
    \includegraphics[width=0.98\linewidth]{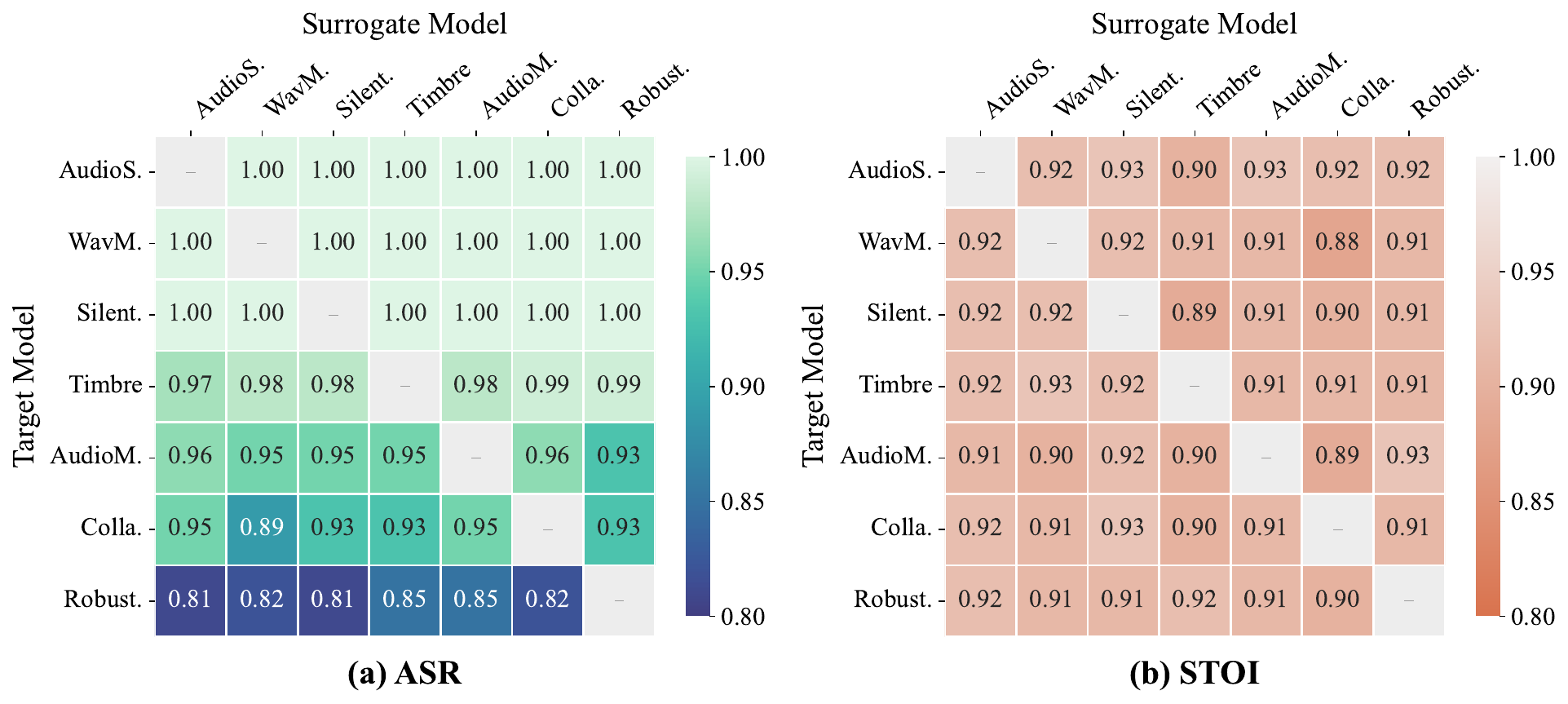}
    \caption{Transferability and fidelity across surrogates.}
    \label{fig:Transferability}
\end{figure}

\emph{\underline{Observation 2}: ScrambleMark transfers consistently regardless of the surrogate choice.} Fig.~\ref{fig:Transferability} shows that for most target watermarking models, ASR remains between 0.93 and 1.00 regardless of which surrogate is used, while STOI stays tightly concentrated around 0.90--0.93. This indicates that the attack does not rely on a particular surrogate watermarking model. Surrogate robustness nevertheless has a mild effect. RobustDNN remains the most difficult target across surrogate choices, likely due to its robustness-oriented training strategy. Weaker surrogates such as WavMark~\cite{chen2023wavmark} also transfer slightly worse to some targets (e.g., 0.89 ASR on Collaborative and 0.82 ASR on RobustDNN), suggesting a greater tendency to capture model-specific artifacts. Overall, these results suggest that ScrambleMark targets shared harmonic structures rather than detector-specific weaknesses.

\subsubsection{Impact of Fine-tuning}\label{sec:ablation}
We examine the impact of fine-tuning. Fig.~\ref{modelalignment} shows that fine-tuning progressively reduces feature drift in unseen speech, while improving STOI. Details are provided in Appendix~\ref{appendix:Academic}.

\subsection{ Evaluations on Commercial Platforms}\label{sec:realworld}
We evaluate ScrambleMark against three commercial systems. We measure evasion effectiveness and assess audio quality using (i) STOI and (ii) human-subject study.

\subsubsection{Commercial Watermarks Setup}\label{commercialsetup}
Commercial speech watermarking comes in two forms, and each dictates how we obtain its watermarked test data.

\textit{(i) Watermarking-as-a-service (WaaS)} lets a user submit their own audio to a service that embeds an imperceptible watermark for ownership protection and leak tracing; Resemble AI~\cite{resemble_watermarker} and TrustedAudio~\cite{trustedaudio_about} offer this model, we build their test sets from English utterances in Section~\ref{Sec:expsetup}: for Resemble AI we use a balanced 3 female / 3 male split and clips of 3 to 12\,s (42 utterances total if matched to the SynthID setup), and for TrustedAudio we use longer clips of around 20\,s to suit its tracing use case (42 utterances total).

\textit{(ii) Generation-native watermarking} embeds the mark into the waveform as a text-to-speech (TTS) model generates speech, certifying content as machine-generated rather than protecting user audio; Google SynthID~\cite{deepmind_synthid} is the representative example. Since its watermark exists only in platform-generated speech, we synthesize the test set with Google's text-to-speech. We use six speakers (3 female and 3 male) and have each read distinct transcripts drawn from LibriSpeech~\cite{panayotov2015librispeech}, producing 42 watermarked utterances of 3 to 12 seconds, with no transcript reused across speakers.

\subsubsection{Objective Speech Quality Evaluation} \label{objectivespeechquality} We first evaluate adversarial audio quality using STOI, enabling large-scale assessment across all attack methods.

\noindent\textbf{Setup.} For WaaS watermarking systems, we adopt the attack configuration described in Section~\ref{Sec:expsetup}. For generation-native watermarking systems, we expand the initial configuration $\boldsymbol{\Omega}^0$ to comprise 200 randomly initialized sub-bands. This warm-start strategy introduces greater diversity into the initial perturbation pattern and enables more effective spectrogram-level modifications. We further evaluate the impact of initial band through a sensitivity analysis in Table~\ref{tab:all_sensitivity}.

\begin{table}[tb]
\centering
\caption{Commercial platforms: ASR \& objective quality}
\setlength{\tabcolsep}{4pt}
\renewcommand{\arraystretch}{0.75} 
\resizebox{0.48\textwidth}{!}{
\begin{tabular}{l cc cc cc| cc}
\toprule
\multirow{2}{*}{Method}
& \multicolumn{2}{c}{SynthID}
& \multicolumn{2}{c}{Resemble AI}
& \multicolumn{2}{c}{TrustedAudio}
& \multicolumn{2}{|c}{\textbf{Average}} \\
\cmidrule(lr){2-3}
\cmidrule(lr){4-5}
\cmidrule(lr){6-7}
\cmidrule(lr){8-9}
& ASR($\uparrow$) & STOI($\uparrow$)
& ASR($\uparrow$) & STOI($\uparrow$)
& ASR($\uparrow$) & STOI($\uparrow$)
& ASR($\uparrow$) & STOI($\uparrow$)\\
\midrule
White-Box-PGD~\cite{liu2024audiomarkbench}
& 0.00 & 1.00 & 0.00 & 1.00 & 0.15 & 1.00 & 0.05 & 1.00 \\
White-Box-I-FGSM~\cite{liu2024audiomarkbench}
& 0.00 & 1.00 & 0.00 & 1.00 & 0.13 & 1.00 & 0.04 & 1.00 \\
Black-Box-HSJA~\cite{liu2024audiomarkbench}
& 0.79 & 0.87 & 1.00 & 0.83 & 1.00 & 0.77 & 0.93 & 0.82 \\
Black-Box-S-HSJA~\cite{liu2024audiomarkbench}
& 0.38 & 0.83 & 1.00 & 0.85 & 1.00 & 0.79 & 0.79 & 0.82 \\
SE Attack~\cite{lopez2024speech}
& 0.00 & 1.00 & 0.00 & 0.99 & 0.00 & 0.99 & 0.00 & 0.99 \\
Overwriting Attack~\cite{yao2026yours}
& 0.00 & 0.98 & 0.00 & 0.98 & 0.00 & 0.99 & 0.00 & 0.98 \\
HarmonicAttack~\cite{li2025harmonicattack}
& 0.00 & 0.95 & 0.00 & 0.98 & 0.00 & 0.97 & 0.00 & 0.97 \\
DiffErase-L~\cite{yaoblack}
& 0.90 & 0.75 & 1.00 & 0.73 & 1.00 & 0.71 & 0.97 & 0.73 \\
DiffErase-M~\cite{yaoblack}
& 0.93 & 0.79 & 1.00 & 0.72 & 1.00 & 0.72 & 0.98 & 0.74 \\
Encodec~\cite{defossez2022encodec}
& 0.90 & 0.78 & 0.98 & 0.73 & 1.00 & 0.69 & 0.96 & 0.73 \\
\midrule
\textbf{ScrambleMark}
& 0.93 & 0.92 & 1.00 & 0.91 & 1.00 & 0.90
& \textbf{0.98} & \textbf{0.91}\\
\bottomrule
\end{tabular}}
\label{tab:commercialevaluation}
\end{table}

\textit{\underline{Observation 3}: ScrambleMark remains effective against deployed commercial watermarking systems.} As shown in Table~\ref{tab:commercialevaluation}, ScrambleMark achieves an average ASR of 0.98 and a STOI of 0.91 across all three platforms. In comparison, the strongest baselines, DiffErase-M and Encodec, achieve similar ASRs (0.98 and 0.96) but substantially lower STOI scores (0.74 and 0.73). As a result, ScrambleMark improves speech intelligibility by approximately 23\% without sacrificing evasion effectiveness. This trend is consistent across all platforms, where ScrambleMark maintains ASRs between 0.93 and 1.00 while preserving high audio quality. 

\subsubsection{Cross-Surrogate Transferability}

We further evaluate whether transferability to commercial platforms depends on the surrogate by testing each academic model.

\begin{table}[h]
\centering
\caption{Cross-Surrogate Transferability and Quality.}
\setlength{\tabcolsep}{4pt}
\renewcommand{\arraystretch}{0.8} 

\resizebox{0.48\textwidth}{!}{
\begin{tabular}{l cc cc cc| cc}
\toprule
\multirow{2}{*}{Surrogate Model}
& \multicolumn{2}{c}{SynthID}
& \multicolumn{2}{c}{Resemble AI}
& \multicolumn{2}{c}{TrustedAudio}
& \multicolumn{2}{|c}{\textbf{Average}} \\
\cmidrule(lr){2-3}
\cmidrule(lr){4-5}
\cmidrule(lr){6-7}
\cmidrule(lr){8-9}
& ASR($\uparrow$) & STOI($\uparrow$)
& ASR($\uparrow$) & STOI($\uparrow$)
& ASR($\uparrow$) & STOI($\uparrow$)
& ASR($\uparrow$) & STOI($\uparrow$)\\
\midrule
AudioSeal
& 0.86 & 0.92 & 0.98 & 0.90 & 1.00 & 0.90 & 0.95 & 0.91 \\
WavMark
& 0.83 & 0.92 & 0.93 & 0.91 & 1.00 & 0.91 & 0.92 & 0.91 \\
AudioMarkNet
& 0.86 & 0.89 & 0.93 & 0.90 & 1.00 & 0.90 & 0.93 & 0.90 \\
SilentCipher
& 0.83 & 0.92 & 0.90 & 0.89 & 1.00 & 0.89 & 0.91 & 0.90 \\
Collaborative
& 0.86 & 0.90 & 0.98 & 0.91 & 1.00 & 0.90 & 0.95 & 0.90 \\
RobustDNN
& 0.90 & 0.90 & 1.00 & 0.91 & 1.00 & 0.91 & 0.97 & 0.91 \\
Timbre
& 0.93 & 0.92 & 1.00 & 0.91 & 1.00 & 0.91 & 0.98 & 0.91 \\

\bottomrule
\end{tabular}}
\label{tab:surrogatecommercial}
\end{table}

\noindent\textbf{Results.} Table~\ref{tab:surrogatecommercial} confirms transferability to commercial platforms. Across surrogates, average ASR is 0.91--0.98 with STOI of 0.90--0.91, reaching 1.00 on TrustedAudio, showing that the transferability extends to commercial platforms.

\subsubsection{Robustness and Sensitivity}
We evaluate ScrambleMark's statistical robustness, parameter sensitivity, robustness to multiple watermarks, and impact on unwatermarked audio. Tables~\ref{tab:statisticalrobustness}, \ref{tab:all_sensitivity}, \ref{tab:multiplewatermarkmatchscore}, and \ref{tab:nowatermark_utility} show stable performance across independent runs, characterize parameter trade-offs, demonstrate effective multi-watermark evasion, and confirm preserved unwatermarked-audio utility, respectively. Results are provided in Appendices~\ref{statisticalrobustness}, \ref{parametersensitivity}, and \ref{multiwatermark}.


\subsubsection{Human-Subject Evaluation}
\label{sec:human_eval}
We conduct a human-subject study to assess the perceptual quality of perturbed speech. To limit participant burden, we consider only attacks that successfully evade commercial systems: DiffErase-L, DiffErase-M, and Black-Box-HSJA. We exclude White-box, SE, Overwriting, and HarmonicAttack due to their low ASR.

\noindent\textbf{Dataset.} 
To cover diverse speech content, we construct a new set comprising 30 samples per platform (5 different utterances from the same 6 speakers set used in Section~\ref{commercialsetup}) to increase utterance diversity and verify ASR stability on new utterances from the same speakers. We evaluate our attack and 3 baselines. Each method generates one adversarial version of every source sample, resulting in 360 adversarial clips across the 3 platforms.

\noindent\textbf{Human Evaluation Setup.}  We recruited 24 normal-hearing university students (12 male, 12 female; aged 20--30) to evaluate the perceived quality of these clips. The study protocol was approved by our Institutional Review Board (IRB). The experiment was conducted in a quiet environment using headphones. Following the standard Mean Opinion Score (MOS) protocol, participants first listened to the original reference audio, followed by the corresponding perturbed speech. They rated the perceptual quality on a continuous 0--5 scale: 0--1 (Bad), 1--2 (Poor), 2--3 (Fair), 3--4 (Good), and 4--5 (Excellent). To prevent cognitive overload, participants were divided into three groups of eight, with each group evaluating 120 clips from a single commercial platform. 

\begin{table}[tb]
\centering
\caption{Commercial platforms: ASR \& human ratings.}
\label{commercialperformance}
\renewcommand{\arraystretch}{0.8} 
\resizebox{0.46\textwidth}{!}{%
\begin{tabular}{l|cc|cc|cc|cc}
\toprule
\multirow{2}{*}{Method}
& \multicolumn{2}{c|}{SynthID}
& \multicolumn{2}{c|}{Resemble AI}
& \multicolumn{2}{c|}{TrustedAudio}
& \multicolumn{2}{c}{Average} \\
& ASR$\uparrow$ & MOS$\uparrow$
& ASR$\uparrow$ & MOS$\uparrow$
& ASR$\uparrow$ & MOS$\uparrow$
& ASR$\uparrow$ & MOS$\uparrow$ \\
\midrule
Black-Box-HSJA~\cite{liu2024audiomarkbench}
& 29/30 & 1.16
& 30/30 & 1.08
& 30/30 & 1.14
& 98.9\% & 1.13 \\

DiffErase-L~\cite{yaoblack}
& 25/30 & 1.52
& 30/30 & 1.35
& 30/30 & 2.29
& 94.4\% & 1.72 \\

DiffErase-M~\cite{yaoblack}
& 28/30 & 1.27
& 30/30 & 1.32
& 30/30 & 2.08
& 97.8\% & 1.56 \\

\textbf{ScrambleMark (Ours)}
& \textbf{27/30} & \cellcolor{green!18}\textbf{3.95}
& \textbf{30/30} & \cellcolor{green!25}\textbf{4.36}
& \textbf{30/30} & \cellcolor{green!23}\textbf{4.20}
& \textbf{96.7\%} & \cellcolor{green!25}\textbf{4.17} \\
\bottomrule
\end{tabular}}
\end{table}

\textit{\underline{Observation 4}: ScrambleMark achieves substantially higher human ratings than existing attacks.} Table~\ref{commercialperformance} reports the MOS of each attack method. ScrambleMark consistently achieves the highest perceptual quality across all platforms, with an average MOS of 4.17 while maintaining successful watermark evasion. In contrast, the strongest baseline, DiffErase-L, achieves an average MOS of only 1.72 despite attaining a comparable attack success rate. Overall, ScrambleMark improves perceptual quality by 2.4 times over DiffErase-L, indicating that effective watermark removal need not come at the expense of audio quality. Appendix~\ref{statisticalsignificance} confirms that ScrambleMark achieves significantly higher MOS ratings than all baselines across three platforms.

\section{Robustness to Adaptive Defenses}\label{sec:robustness}

\subsection{Adaptive Recovery Design}
We consider a defender that suspects a speech has been scrambled and tries to recover it. To grant the defender a strong possible footing, we assume it holds the original watermarked audio $x_{w_t}$ as a reference. The defender applies the same band-wise operator $\mathcal{H}$ used by the adversary, which performs a per-band frequency shift $\mathbf{S}$ over a band partition $\boldsymbol{\Omega}$, to map the scrambled speech $x'_s$ to a recovered audio $x_{rec} = \mathcal{H}(x'_s \mid (\boldsymbol{\Omega}, \mathbf{S}))$. The defender optimizes $\mathbf{S}$ so that $x_{rec}$ approaches the reference $x_{w_t}$ under two objectives.

\noindent\textit{(i) Perceptual recovery} drives $x_{rec}$ back toward $x_{w_t}$ in perceptual quality, under a Mel loss:
\begin{equation}
    \min_{\mathbf{S}}\; \mathcal{L}_{\text{Mel}}(x_{rec},\, x_{w_t}).
    \label{recoverpreception}
\end{equation}

\noindent\textit{(ii) Bit recovery} drives the bits extracted from $x_{rec}$ back toward those of $x_{w_t}$, maximizing the matching score $\Theta$:
\begin{equation}
    \max_{\mathbf{S}}\; \Theta(\mathcal{D}(x_{rec}),\, \mathcal{D}(x_{w_t})).
    \label{accrecover}
\end{equation}

\noindent\textbf{Recovery Procedure.} We assume an adaptive defender that knows the attack algorithm and attempts to recover the per-band shifts applied by the attacker. The spectrogram is partitioned into 500 equal-width frequency bands, and the shift vector $\mathbf{S}$ is optimized via greedy search. For each band, candidate shifts are evaluated at 15-cent intervals, and the best shift is retained. The procedure is repeated for five passes over all bands with early stopping upon convergence.



\subsection{Robustness Evaluation}

\noindent\textbf{Setup.}
To assess recovery robustness, we compare ScrambleMark with the deterministic scrambler in two settings. \textit{(i) Academic schemes}: we select 350 successful ScrambleMark samples per scheme from Section~\ref{comparisonattack}, totaling 2,100 samples across six schemes. \textit{(ii) Commercial platforms}: we randomly select 5 successful samples per speaker from the six speakers in Sections~\ref{objectivespeechquality} and~\ref{sec:human_eval}, resulting in 90 samples across three platforms. Deterministic samples are generated using Algorithm~\ref{alg:BA} under the configuration of Section~\ref{sec:al-evasionsetup} and retained only if they evade the corresponding detector, matching ScrambleMark in sample count.

Since commercial platforms expose only binary detection outcomes and no matching scores, we evaluate perceptual recovery only. We further exclude RobustDNN from bit recovery because optimizing for bit recovery inflates its false positive rate (FPR), causing speech carrying non-target messages to be identified as the target watermark. Detailed FPR analysis for each model can be found in Appendix~\ref{appendix:fpr}.

\noindent\textbf{Metric.} We evaluate \textit{pre-recovery} and \textit{post-recovery ASR}, corresponding to attack success before and after recovery, respectively. The drop between the two reflects the attack's robustness to recovery: smaller drops indicate that more attack effectiveness is retained after recovery.

\begin{figure}
    \centering
    \includegraphics[width=0.90\linewidth]{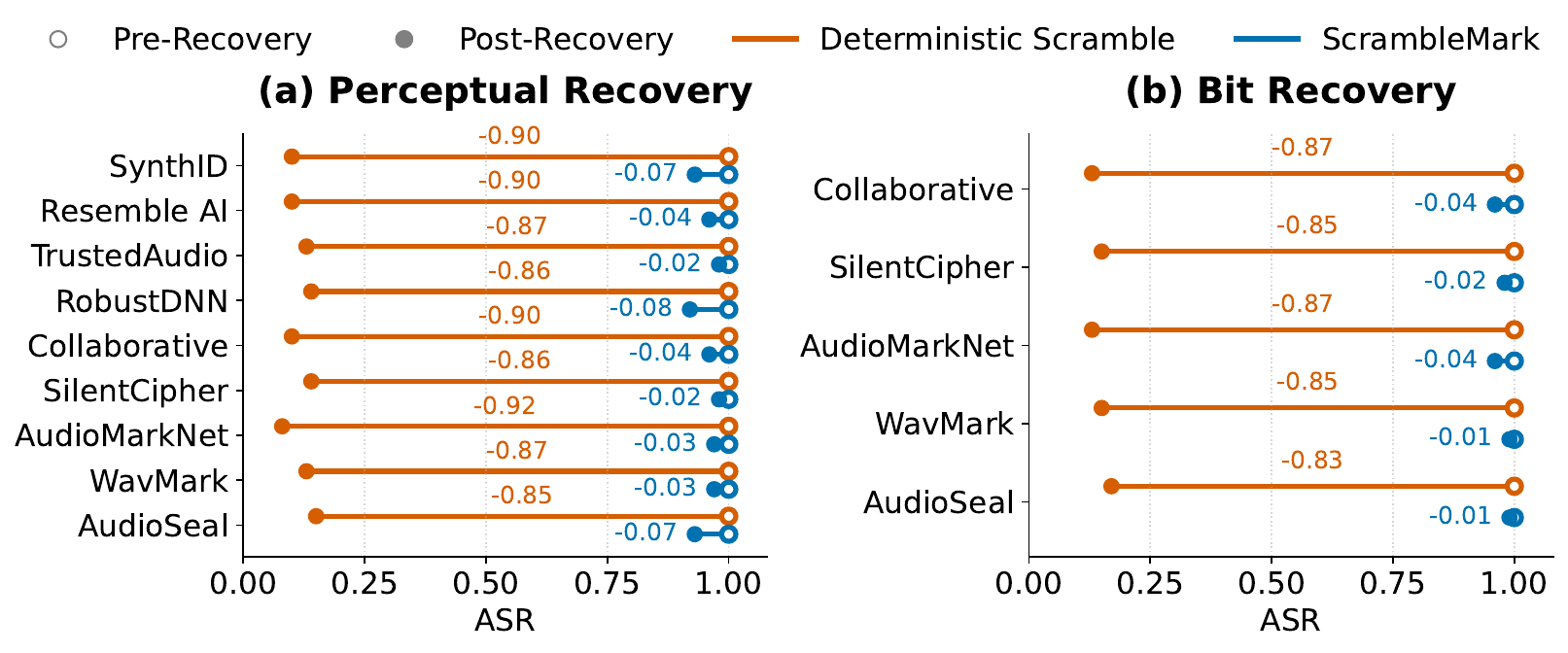}
    \caption{Attack robustness under recovery.}
    \label{recover_analysis}
\end{figure}

\textit{\underline{Observation 5}: ScrambleMark remains robust against adaptive recovery.} Fig.~\ref{recover_analysis} reports ASR before and after perceptual and bit recovery. On academic watermarking schemes, recovery reduces the deterministic scrambler's ASR by 0.88 on average, whereas ScrambleMark incurs only a 0.04 drop. On commercial platforms, the corresponding drops are 0.85 and 0.02. In both settings, recovery has only a marginal impact on ScrambleMark's effectiveness. This robustness stems from its learned scrambling distribution, which does not expose a deterministic recovery path.

Our adaptive recovery focuses on reversing the known band-wise shift operator. A stronger defender may train a detector-aware recovery model and search over intensity levels using detector feedback. While feasible, such recovery remains non-trivial because band shifting may merge or split features, causing information loss and ambiguous inversion. Reconstructing missing features may further remove watermark evidence or introduce watermark-like artifacts. These stronger recovery strategies are worth further study.

\noindent\textbf{Ethical Considerations:}
Vulnerability disclosure and human-subject considerations are discussed in Appendix~\ref{sec:ethic}.

\section{Related Work} 

\noindent\textbf{Image Watermarking Evasion Attacks.} Watermark evasion has been studied more extensively for images, spanning white-box, black-box, and no-box settings. These attacks share a common recipe: they locate the watermark in a fixed representation, such as the image spectrum~\cite{kassis2025unmarker}, the bit distribution~\cite{jiang2023evading}, or a VAE latent space~\cite{lin2025crack,hutransfer}, and then perturb that representation to remove the watermark until detection fails. Our ScrambleMark uses harmonic scrambling to disrupt watermark integrity without localization.

\noindent\textbf{Adversarial Speech Attacks.} Existing speech attacks add perturbations that target speech recognition~\cite{carlini2018audio,liu2022evil,wang2020towards,cheng2024alif,fang2024zero,wu2025translators} or speaker recognition~\cite{chen2021real,li2020advpulse,li2023inaudible,yusmack,duanparrot}. Because their perturbation only needs to flip the model, it stays small and localized. Full-spectrum watermarks require degrading full-spectrum perturbations, whereas ScrambleMark preserves naturalness by scrambling harmonics.

\noindent\textbf{Generative Models for Speech.} Generative speech models such as voice cloning and text-to-speech~\cite{du2024cosyvoice,eskimez2024e2tts,wang2023valle} are built to synthesize a target speaker's timbre, reproducing who is speaking. None of them is designed to scramble the harmonic structure of audio, which is what watermark evasion requires. Repurposing these models for evasion is also not straightforward: trained to reconstruct or imitate seen voices, they distort audio drawn from unseen distributions~\cite{yaoblack}, the same failure we observe for unsupervised removal attacks. ScrambleMark adapts a generative model to perform harmonic scrambling and, through its pretraining and fine-tuning design, holds this behavior on unseen inputs.

\section{Conclusion} 
We introduce \emph{structural evasion}, a new paradigm showing that speech watermarks can be evaded without being removed. Building on this insight, we propose ScrambleMark, a black-box harmonic-scrambling method that achieves high evasion across ten watermark models while preserving speech quality and resisting adaptive recovery.

\section*{Acknowledgments}
This work was supported in part by the U.S. National Science Foundation (NSF) under Grants CNS-2451168, CCF-2551963, DGE-2623286, and CCF-2551964.

\bibliographystyle{IEEEtran}
\bibliography{reference}


\appendices

\begin{figure*}[t]
\centering
\includegraphics[width=1\linewidth]{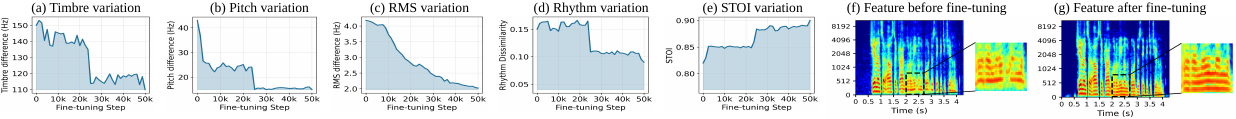}
\caption{Perceptual feature variation across fine-tuning steps, and harmonic features before and after fine-tuning.}
\label{modelalignment}
\end{figure*}

\section{Feature-variation in Generative Models}\label{feature}

We evaluate feature changes on seen and unseen speech using RAVE~\cite{caillon2021rave}, Mel-VAE~\cite{liu2023audioldm}, HiFi-GAN~\cite{kong2020hifi}, BigVGAN~\cite{lee2022bigvgan}, FastDiff~\cite{huang2022fastdiff}, and DiffWave~\cite{kong2021diffwave}.

\noindent\textbf{Datasets.} All models are trained on LJ Speech and evaluated on VoxCeleb1. We randomly sample 500 utterances from each dataset as \emph{seen} and \emph{unseen} data, respectively, and reconstruct each utterance using the corresponding model.

\noindent\textbf{Metrics.}  We measure output--input variations across four perceptual features using RMS difference~\cite{skovenborg2004evaluation}, $F_0$ difference~\cite{jadoul2018introducing,boersma1993accurate}, MCD~\cite{kubichek1993mel,toda2007voice}, and EED~\cite{foote2002audio}. Following prior work~\cite{chhaglani2025artifree,jiang2025refvc}, all metrics are normalized to $[0,1]$, with lower values indicating smaller variations.

\section{Academic Models Evaluation}\label{appendix:Academic}

\noindent\textbf{Threshold Evaluation.}
We use each scheme's default threshold $\kappa$; for Collaborative and RobustDNN, we balance FPR and false negative rate (FNR) following~\cite{liu2024audiomarkbench}. Table~\ref{tab:threshold} reports the thresholds and low FNRs.

\begin{table}[h]
\centering
\caption{Detection threshold $\kappa$ for different watermarks.}
\label{Threshold}
\resizebox{\columnwidth}{!}{
\begin{tabular}{lccccccc}
\specialrule{0.10em}{0pt}{0pt}  
\textbf{Metric} & \textbf{Timbre} & \textbf{AudioSeal} & \textbf{WavMark} & \textbf{AudioMarkNet} & \textbf{SilentCipher} & \textbf{Collaborative} & \textbf{RobustDNN} \\
\midrule
Threshold $\kappa$ & 0.90 & 0.50 & 0.50 & 0.99 & 0.50 & 0.50 & 0.99 \\
FPR & 0.00 & 0.00 & 0.00 & 0.00 & 0.00 & 0.02 & 0.00 \\
FNR & 0.00 & 0.00 & 0.00 & 0.00 & 0.03 & 0.00 & 0.06 \\
\specialrule{0.10em}{0pt}{0pt}
\end{tabular}
}\label{tab:threshold}
\end{table}

\noindent\textbf{Impact of Fine-Tuning.} We analyze perceptual changes during fine-tuning using the same 500 English samples in Table~\ref{tab:existattackevaluation}. We evaluate four perceptual features using $F_0$, MCD, RMS, and EED differences, together with STOI.

\noindent\textbf{Results.}
Fig.~\ref{modelalignment} (a), (b), (c), and (d) show that all feature deviations decrease during fine-tuning, while STOI improves from 0.82 to 0.90, with most gains occurring within 20k steps. Fig.~\ref{modelalignment} (f) and (g) visualize the spectrogram changes.

\noindent\textbf{Computation Overhead.} 
We evaluate ScrambleMark at inference time on an NVIDIA RTX 4090 GPU and an Intel Core i9-14900K CPU, using DiffWave fast sampling with six sampling steps. The ScrambleMark model occupies 422 MiB of GPU memory and 546 MiB of system RAM, while the peak GPU memory usage reaches 1,174 MiB for a single input. Generating one candidate takes 0.41 s per candidate and about 5 s for ten-candidate generation and selection.


\section{False Positive Rate Validation of Recovery} \label{appendix:fpr}





A recovery objective should maintain a low FPR to avoid falsely attributing non-target speech to the target watermark.
\noindent\textbf{Setup.}
For each scheme, we randomly select 500 clean samples and generate target and non-target watermarked versions. We apply each recovery objective using 500 equal-width frequency bands and report the resulting FPR in Table~\ref{FPR}. Collaborative supports only binary watermark-presence detection and thus cannot evaluate non-target messages; its entries are marked as ``--''.

\begin{table}[h]
\centering
\renewcommand{\arraystretch}{0.65}
\caption{FPR across recovery objectives.}
\label{FPR}
\resizebox{\columnwidth}{!}{
\begin{tabular}{ccccccccc}
\specialrule{0.10em}{0pt}{0pt}
\textbf{Objective} & Type & \textbf{AudioS.} & \textbf{WavM.} & \textbf{AudioM.} & \textbf{Silent.} & \textbf{Colla.} & \textbf{Robust.} \\
\midrule
\multirow{2}{*}{Percp. Recovery} & clean & 0.00 & 0.00 & 0.00 & 0.00 & 0.00 & 0.00 \\
& untarget & 0.00 & 0.00 & 0.00 & 0.00 & - & 0.00 \\
\multirow{2}{*}{Bit Recovery} & clean & 0.00 & 0.00 & 0.00 & 0.00 & 0.00 & 1.00 \\
& untarget & 0.00 & 0.00 & 0.00 & 0.00 & - & 1.00 \\
\specialrule{0.10em}{0pt}{0pt}
\end{tabular}
}
\end{table}

\noindent\textbf{Results.} Table~\ref{FPR} shows near-zero FPR for perceptual recovery across all schemes, while bit recovery yields 1.00 FPR on RobustDNN. We therefore use perceptual recovery by default and bit recovery only when it maintains low FPR.

\section{Robustness and Sensitivity.} \label{robustness_sensitivity}

\subsection{Statistical Robustness}\label{statisticalrobustness}

We evaluate the stability of ScrambleMark's performance under random factors by independently repeating the entire pipeline 15 times, while fixing the evaluation samples and deterministic configurations across runs. We use the test sets from Tables~\ref{tab:existattackevaluation} and~\ref{tab:commercialevaluation} for academic and commercial models, respectively.

\begin{table}[h]
\centering
\renewcommand{\arraystretch}{0.85}
\caption{ASR and STOI statistics over 15 runs.}
\setlength{\tabcolsep}{5pt}
\renewcommand{\arraystretch}{1.15}

\resizebox{0.48\textwidth}{!}{
\begin{tabular}{ll cccccccccc}
\toprule
Metric & Statistic
& AudioSeal & WavMark & AudioMark. & SilentC.
& Collab. & Robust. & Timbre* & SynthID & Resemble & TrustedA. \\
\midrule

\multirow{2}{*}{ASR}
& Mean
& 1.00 & 1.00 & 0.98 & 1.00 & 0.94 & 0.87 & 1.00  & 0.90 & 0.98 & 1.00\\
& Variance
& 0.0000 & 0.0000 & 0.0002 & 0.0000 & 0.0005 & 0.0005 & 0.0000 & 0.0007 & 0.0003 & 0.0000\\

\midrule

\multirow{2}{*}{STOI}
& Mean
& 0.90 & 0.91 & 0.89 & 0.91 & 0.90 & 0.89 & 0.92 & 0.91 & 0.90 & 0.90\\
& Variance
& 0.0001 & 0.0002 & 0.0002 & 0.0003 & 0.0002 & 0.0003 & 0.0003 & 0.0002 & 0.0002 & 0.0003\\

\bottomrule
\end{tabular}}
\label{tab:statisticalrobustness}
\end{table}

\noindent \textbf{Results.} As shown in Table~\ref{tab:statisticalrobustness}, ScrambleMark achieves mean ASRs of 0.87--1.00 across academic models and commercial platforms, while the ASR variance remains below 0.001 in all cases. The mean STOI scores range from 0.89 to 0.92, with even smaller variances of 0.0001--0.0003. Overall, ScrambleMark maintains stable performance across independent runs of its randomized pipeline.

\subsection{Detector Threshold Sensitivity}\label{thresholdsensitivity}
Beyond the fixed thresholds in Table~\ref{tab:existattackevaluation}, we evaluate ScrambleMark across different thresholds to assess sensitivity. One way to make the detector more challenging for ScrambleMark is to increase the FPR by lowering the detection threshold. As more clean audio is classified as watermarked, ScrambleMark-processed audio also becomes more likely to be detected as watermarked. We progressively lower each detector's threshold by reducing the required matching bits to obtain the FPR ranges in Table~\ref{tab:thresholdsensitivity}. The original thresholds in Table~\ref{tab:threshold} serve as the baseline.


\begin{table}[h]
\centering
\renewcommand{\arraystretch}{0.75}
\caption{Threshold sensitivity across FPR levels.}
\label{tab:threshold_sensitivity}

\resizebox{0.47\textwidth}{!}{
\begin{tabular}{c|cc|cc|cc|cc|cc|cc|cc|c}
\toprule
\multirow{2}{*}{\textbf{FPR}}
& \multicolumn{2}{c|}{\textbf{AudioSeal}}
& \multicolumn{2}{c|}{\textbf{AudioMarkNet}}
& \multicolumn{2}{c|}{\textbf{SilentCipher}}
& \multicolumn{2}{c|}{\textbf{Collaborative}}
& \multicolumn{2}{c|}{\textbf{RobustDNN}}
& \multicolumn{2}{c|}{\textbf{Timbre*}}
& \textbf{Avg.} \\

& $\boldsymbol{\kappa}$ & \textbf{ASR}
& $\boldsymbol{\kappa}$ & \textbf{ASR}
& $\boldsymbol{\kappa}$ & \textbf{ASR}
& $\boldsymbol{\kappa}$ & \textbf{ASR}
& $\boldsymbol{\kappa}$ & \textbf{ASR}
& $\boldsymbol{\kappa}$ & \textbf{ASR}
& \textbf{ASR} \\
\midrule
Baseline & 0.500 & 1.00 & 0.990 & 1.00 & 0.500 & 1.00 & 0.500 & 0.96 & 0.990 & 0.86 & 0.900 & 1.00 & 0.97 \\
1-2\% & 0.100 & 1.00 & 0.688 & 0.97 & 0.475 & 0.97 & 0.500 & 0.96 & 0.984 & 0.86 & 0.700 & 0.99 & 0.96 \\
2-5\% & 0.050 & 0.98 & 0.625 & 0.95 & 0.450 & 0.94 & 0.107 & 0.95 & 0.977 & 0.84 & 0.600  & 0.95 & 0.94 \\
5-8\% & 0.020 & 0.92 & 0.563 & 0.91 & 0.425 & 0.93 & 0.004 & 0.91 & 0.965 & 0.82 & 0.500  & 0.93 & 0.90 \\
8-15\% & 0.010 & 0.92 & 0.500 & 0.90 & 0.400 & 0.92 & 0.001 & 0.91 & 0.953 & 0.80 & 0.400  & 0.92 & 0.90 \\

\bottomrule
\end{tabular}
}
\label{tab:thresholdsensitivity}
\end{table}

\begin{table*}[t]
\centering
\renewcommand{\arraystretch}{0.65}
\caption{Sensitivity analysis of ScrambleMark parameters.}
\label{tab:all_sensitivity}
\setlength{\tabcolsep}{2.2pt}

\resizebox{\textwidth}{!}{
\begin{tabular}{
l|cc|cc|cc|cc
||cc|cc|cc|cc
||cc|cc|cc|cc
||cc|cc|cc|cc
}
\toprule

\multirow{3}{*}{\textbf{Target}}
& \multicolumn{8}{c||}{\textbf{Evasion–quality Trade-off Weight}}
& \multicolumn{8}{c||}{\textbf{Diffusion Schedule Steps}}
& \multicolumn{8}{c||}{\textbf{Candidate-Count}}
& \multicolumn{8}{c}{\textbf{Initial-Band-Count}}

\\
\cmidrule(lr){2-9}
\cmidrule(lr){10-17}
\cmidrule(lr){18-25}
\cmidrule(lr){26-33}
& \multicolumn{2}{c|}{\textbf{0.25}}
& \multicolumn{2}{c|}{\textbf{0.5}}
& \multicolumn{2}{c|}{\textbf{1}}
& \multicolumn{2}{c||}{\textbf{2}}

& \multicolumn{2}{c|}{\textbf{30}}
& \multicolumn{2}{c|}{\textbf{40}}
& \multicolumn{2}{c|}{\textbf{50}}
& \multicolumn{2}{c||}{\textbf{70}}

& \multicolumn{2}{c|}{\textbf{1}}
& \multicolumn{2}{c|}{\textbf{3}}
& \multicolumn{2}{c|}{\textbf{5}}
& \multicolumn{2}{c||}{\textbf{10}}

& \multicolumn{2}{c|}{\textbf{1}}
& \multicolumn{2}{c|}{\textbf{50}}
& \multicolumn{2}{c|}{\textbf{100}}
& \multicolumn{2}{c}{\textbf{200}}
\\

& \textbf{ASR} & \textbf{STOI}
& \textbf{ASR} & \textbf{STOI}
& \textbf{ASR} & \textbf{STOI}
& \textbf{ASR} & \textbf{STOI}

& \textbf{ASR} & \textbf{STOI}
& \textbf{ASR} & \textbf{STOI}
& \textbf{ASR} & \textbf{STOI}
& \textbf{ASR} & \textbf{STOI}

& \textbf{ASR} & \textbf{STOI}
& \textbf{ASR} & \textbf{STOI}
& \textbf{ASR} & \textbf{STOI}
& \textbf{ASR} & \textbf{STOI}

& \textbf{ASR} & \textbf{STOI}
& \textbf{ASR} & \textbf{STOI}
& \textbf{ASR} & \textbf{STOI}
& \textbf{ASR} & \textbf{STOI}
\\
\midrule

AudioSeal
& 1.00 & 0.81 & 1.00 & 0.85 & 1.00 & 0.91 & 0.93 & 0.93
& 0.96 & 0.86 & 0.98 & 0.88 & 1.00 & 0.91 & 1.00 & 0.90
& 0.91 & 0.80 & 0.93 & 0.82 & 0.97 & 0.85 & 1.00 & 0.91
& 1.00 & 0.91 & 0.99 & 0.91 & 1.00 & 0.89 & 1.00 & 0.89
\\

AudioMarkNet
& 1.00 & 0.80 & 0.98 & 0.83 & 0.98 & 0.88 & 0.90 & 0.94
& 0.95 & 0.86 & 0.97 & 0.87 & 0.98 & 0.89 & 0.98 & 0.90
& 0.85 & 0.81 & 0.89 & 0.83 & 0.94 & 0.83 & 0.98 & 0.88
& 0.98 & 0.90 & 0.95 & 0.90 & 0.98 & 0.92 & 0.98 & 0.91
\\

SilentCipher
& 1.00 & 0.81 & 1.00 & 0.86 & 1.00 & 0.92 & 0.91 & 0.94
& 0.97 & 0.87 & 0.99 & 0.89 & 1.00 & 0.92 & 1.00 & 0.91
& 0.92 & 0.80 & 0.93 & 0.81 & 0.97 & 0.86 & 1.00 & 0.92
& 0.99 & 0.92 & 0.97 & 0.91 & 0.97 & 0.92 & 1.00 & 0.90
\\

Collaborative
& 1.00 & 0.82 & 0.96 & 0.85 & 0.96 & 0.90 & 0.92 & 0.93
& 0.90 & 0.86 & 0.93 & 0.86 & 0.96 & 0.90 & 0.97 & 0.91
& 0.88 & 0.83 & 0.90 & 0.83 & 0.92 & 0.87 & 0.96 & 0.90
& 0.97 & 0.91 & 0.96 & 0.89 & 0.97 & 0.92 & 0.98 & 0.91
\\

RobustDNN
& 0.95 & 0.80 & 0.90 & 0.83 & 0.86 & 0.93 & 0.80 & 0.94
& 0.84 & 0.85 & 0.84 & 0.87 & 0.86 & 0.93 & 0.89 & 0.93
& 0.79 & 0.83 & 0.81 & 0.85 & 0.82 & 0.88 & 0.86 & 0.93
& 0.89 & 0.88 & 0.86 & 0.90 & 0.86 & 0.92 & 0.88 & 0.90
\\

Timbre*
& 1.00 & 0.81 & 1.00 & 0.85 & 1.00 & 0.91 & 0.93 & 0.95
& 0.97 & 0.87 & 0.99 & 0.88 & 1.00 & 0.91 & 1.00 & 0.92
& 0.89 & 0.82 & 0.92 & 0.85 & 0.97 & 0.87 & 1.00 & 0.91
& 0.99 & 0.90 & 0.97 & 0.91 & 1.00 & 0.90 & 1.00 & 0.89
\\

\midrule

SynthID
& 0.95 & 0.81 & 0.93 & 0.83 & 0.93 & 0.92 & 0.81 & 0.94
& 0.90 & 0.86 & 0.88 & 0.85 & 0.93 & 0.92 & 0.93 & 0.93
& 0.78 & 0.81 & 0.80 & 0.84 & 0.89 & 0.88 & 0.93 & 0.92
& 0.87 & 0.91 & 0.90 & 0.90 & 0.87 & 0.90 & 0.90 & 0.91
\\

ResembleAI
& 1.00 & 0.83 & 1.00 & 0.85 & 1.00 & 0.91 & 0.93 & 0.94
& 0.93 & 0.85 & 1.00 & 0.87 & 1.00 & 0.91 & 1.00 & 0.93
& 0.88 & 0.82 & 0.90 & 0.83 & 0.95 & 0.87 & 1.00 & 0.91
& 0.97 & 0.91 & 0.98 & 0.92 & 0.97 & 0.91 & 1.00 & 0.90
\\

TrustedAudio
& 1.00 & 0.80 & 1.00 & 0.82 & 1.00 & 0.90 & 0.95 & 0.93
& 1.00 & 0.86 & 1.00 & 0.87 & 1.00 & 0.90 & 1.00 & 0.92
& 0.90 & 0.80 & 0.95 & 0.92 & 0.98 & 0.88 & 1.00 & 0.90
& 1.00 & 0.90 & 1.00 & 0.92 & 1.00 & 0.88 & 1.00 & 0.90
\\

\midrule

\textbf{Average}
& \textbf{0.99} & \textbf{0.81}
& \textbf{0.97} & \textbf{0.84}
& \textbf{0.97} & \textbf{0.91}
& \textbf{0.90} & \textbf{0.94}

& \textbf{0.94} & \textbf{0.86}
& \textbf{0.95} & \textbf{0.87}
& \textbf{0.97} & \textbf{0.91}
& \textbf{0.97} & \textbf{0.92}

& \textbf{0.87} & \textbf{0.81}
& \textbf{0.89} & \textbf{0.84}
& \textbf{0.93} & \textbf{0.87}
& \textbf{0.97} & \textbf{0.91}

& \textbf{0.96} & \textbf{0.90}
& \textbf{0.95} & \textbf{0.91}
& \textbf{0.96} & \textbf{0.91}
& \textbf{0.97} & \textbf{0.90}
\\

\bottomrule
\end{tabular}
}
\end{table*}

\noindent\textbf{Results.}
As shown in Table~\ref{tab:thresholdsensitivity}, relaxing the detection thresholds makes watermark detection more permissive and gradually reduces the average ASR from 0.97 to 0.90. Even at the highest FPR range (8--15\%), ScrambleMark still achieves 0.80--0.92 ASR across detectors. For WavMark, the FPR remains 0 even at a zero threshold because it returns no matching score when no watermark is detected~\cite{liu2024audiomarkbench}; ScrambleMark maintains 1.00 ASR in this case.

\subsection{ScrambleMark Parameters} \label{parametersensitivity}

We evaluate ScrambleMark's sensitivity to four parameters: (i) the evasion--quality weight ($\lambda={0.25, 0.5, 1, 2}$), (ii) diffusion schedule steps (30, 40, 50, and 70), (iii) candidate count ($N_\ell={1, 3, 5, 10}$), and (iv) initial band count (1, 50, 100, and 200). We use the same academic and commercial test sets as Tables~\ref{tab:existattackevaluation} and~\ref{tab:commercialevaluation}. For (iii), the 3- and 5-candidate settings use levels approximately spanning the full intensity range. The 1-candidate setting cycles through all ten.

\noindent\textbf{Results.}
We summarize the key observations from Table~\ref{tab:all_sensitivity}: (i) Increasing $\lambda$ improves quality but reduces evasion; $\lambda=1$ provides a strong balance (0.97 ASR/0.91 STOI), while $\lambda=0.25$ achieves 0.99/0.81 and $\lambda=2$ achieves 0.90/0.94. (ii) Increasing diffusion steps improves performance up to 50 steps (0.97/0.91), with negligible gains at 70 (0.97/0.92). (iii) More candidates consistently improve both ASR and STOI, from 0.87/0.81 with one candidate to 0.97/0.91 with ten. (iv) Initial band has little impact (0.95--0.97 average ASR), as optimization converges to approximately 250-300 bands across different initializations, beyond which further splitting minimally reduces the objective loss. Overall, these results characterize the trade-offs of our default settings.

\subsection{Multi- and No-Watermark Evaluation} \label{multiwatermark}

\noindent\textbf{Embedding-Order Study.} Multiple watermarks may improve robustness against evasion, but sequential embedding can introduce overwriting~\cite{yao2026yours}, where later watermarks interfere with earlier ones. We compare two strategies: \textit{Robust-First (RF)}, which places more robust watermarks earlier, and \textit{Weak-First (WF)}, which places less robust ones earlier. We construct three orders for each strategy by varying the order within each robustness group. Using SynthID-watermarked samples from Table~\ref{tab:commercialevaluation} as the base, we sequentially embed the remaining nine watermarks; Table~\ref{tab:multiplewatermarkmatchscore} lists each order from first to last. Each setting is repeated 10 times with different messages. We report matching scores only for academic models, as commercial scores are inaccessible.

\begin{table}[h]
\renewcommand{\arraystretch}{0.55}
\centering
\caption{ Matching scores across embedding orders.}
\label{tab:multiplewatermarkmatchscore}

\setlength{\tabcolsep}{2.5pt}
\renewcommand{\arraystretch}{0.9}
\resizebox{0.47\textwidth}{!}{
\begin{tabular}{c c|ccccccc|c}
\toprule
\textbf{Setting} & \textbf{Order (first $\rightarrow$ last)}
& \textbf{10} & \textbf{9} & \textbf{8}
& \textbf{7} & \textbf{4} & \textbf{5} & \textbf{3}
& \textbf{Avg.} \\
\midrule

Baseline & --
& 1.000 & 1.000 & 1.000 & 1.000
& 1.000 & 1.000 & 1.000 & 1.000 \\

\midrule

RF-I & 1-2-4-3-5-7-6-10-8-9
& 0.735 & 1.000 & 0.958 & 0.727
& 0.994 & 0.959& 0.991 & 0.909 \\ 

RF-II & 1-2-5-4-3-6-7-9-10-8
& 0.952 & 1.000 & 1.000 & 0.807 
& 0.995 & 0.931 & 0.995 & 0.954 \\

RF-III & 1-2-3-5-4-7-6-9-8-10
& 1.000 & 1.000 & 0.961 & 0.752 
& 0.997 & 0.946& 0.990 & 0.949 \\

\midrule

WF-I & 1-2-9-10-8-6-7-5-4-3
& 0.411 & 0.450 & 0.603 & 0.709
& 0.997 & 0.977 & 1.000 & 0.735 \\

WF-II & 1-2-8-9-10-7-6-4-3-5
& 0.479 & 0.495 & 0.579 & 0.704 
& 0.995 & 1.000 & 0.993 & 0.749 \\

WF-III & 1-2-10-8-9-6-7-3-5-4
& 0.395 & 0.538 & 0.582 & 0.713 
& 1.000 & 0.983 & 0.992 & 0.743 \\

\bottomrule
\end{tabular}
}

\vspace{1mm}
{\scriptsize
1: SynthID; 2: Resemble AI; 3: AudioMarkNet; 4: RobustDNN; 5: Timbre;\\[-2pt]
6: TrustedAudio; 7: Collaborative; 8: AudioSeal; 9: WavMark; 10: SilentCipher.
}

\end{table}

\noindent\textbf{Results.} Table~\ref{tab:multiplewatermarkmatchscore} shows clear overwriting effects across sequentially embedded watermarks. RF preserves substantially higher average matching scores (0.909--0.954) than WF (0.735--0.749), as weaker watermarks are more vulnerable to subsequent overwriting.


\noindent\textbf{Multi-Watermark Evaluation.}
We evaluate ScrambleMark against RF-(i)--(iii), with WF-(i)--(iii) and Random as baselines, using the same 42 SynthID samples from Table~\ref{tab:commercialevaluation}. Each watermark is embedded and verified as detectable before attack; we define All-Watermark ASR (AW-ASR) as the fraction of samples that simultaneously evade all ten detectors and Single-Watermark ASR (SW-ASR) is the fraction of multi-watermarked samples that evade an individual embedded watermark. Each setting is repeated five times.

\begin{table}[h]
\centering
\caption{ScrambleMark against multiple watermarks.}
\label{tab:multiplewatermarkeffectiveness}
\setlength{\tabcolsep}{3pt}
\renewcommand{\arraystretch}{0.9}

\resizebox{0.47\textwidth}{!}{
\begin{tabular}{llcccccccccc}
\toprule


\textbf{Setting} & \textbf{Metric}
& \textbf{AudioSeal}
& \textbf{WavMark}
& \textbf{AudioMark.}
& \textbf{SilentC.}
& \textbf{Collab.}
& \textbf{Robust.}
& \textbf{Timbre}
& \textbf{SynthID}
& \textbf{Resemble}
& \textbf{TrustedA.} \\
\midrule



RF
& SW-ASR
& 1.00 & 1.00 & 1.00 & 1.00 & 1.00
& 0.91 & 1.00 & 0.93 & 1.00 & 1.00 \\

WF
& SW-ASR
& 1.00 & 1.00 & 0.99 & 1.00 & 1.00
& 0.96 & 1.00 & 0.96 & 1.00 & 1.00 \\

Random
& SW-ASR
& 1.00 & 1.00 & 1.00 & 1.00 & 1.00
& 0.95 & 1.00 & 0.95 & 1.00 & 1.00 \\

\midrule

\multicolumn{12}{c}{\textbf{Overall Multi-Watermark Performance}} \\
\midrule

\multicolumn{4}{c}{\textbf{RF}}
& \multicolumn{4}{c}{\textbf{WF}}
& \multicolumn{4}{c}{\textbf{Random}} \\

\cmidrule(lr){1-4}
\cmidrule(lr){5-8}
\cmidrule(lr){9-12}

\multicolumn{2}{c}{\textbf{AW-ASR}}
& \multicolumn{2}{c}{\textbf{STOI}}
& \multicolumn{2}{c}{\textbf{AW-ASR}}
& \multicolumn{2}{c}{\textbf{STOI}}
& \multicolumn{2}{c}{\textbf{AW-ASR}}
& \multicolumn{2}{c}{\textbf{STOI}} \\

\multicolumn{2}{c}{0.90}
& \multicolumn{2}{c}{0.91}
& \multicolumn{2}{c}{0.95}
& \multicolumn{2}{c}{0.90}
& \multicolumn{2}{c}{0.92}
& \multicolumn{2}{c}{0.90} \\

\bottomrule
\end{tabular}}
\end{table}











\noindent\textbf{Results.} ScrambleMark achieves AW-ASRs of 0.90, 0.95, and 0.92 under RF, WF, and Random, respectively, with RF providing the strongest defense. Multi-watermark embedding can also weaken individual watermarks through overwriting; for example, SynthID's SW-ASR increases from 0.91 in the single-watermark baseline (Table~\ref{tab:statisticalrobustness}) to 0.93--0.96 across the three multi-watermark embedding orders. These results demonstrate the benefit of structural evasion in transferring across diverse watermark configurations.

\noindent\textbf{No-Watermark.} 
We further evaluate whether applying ScrambleMark to unwatermarked audio unnecessarily degrades speech utility.
We randomly select 200 clean samples from each of the five languages in Section~\ref{Sec:expsetup}, all verified as unwatermarked by the evaluated detectors, and apply ScrambleMark. We measure STOI, speaker preservation rate (SPR) using SpeechBrain~\cite{ravanelli2021speechbrain}, and word error rate (WER) using Whisper large-v3~\cite{radford2023robust}. SPR measures the fraction of outputs preserving the original speaker, while WER is computed between clean and perturbed audio transcripts.

\begin{table}[h]
\centering
\renewcommand{\arraystretch}{0.55}
\caption{Speech utility on unwatermarked audio.}
\label{tab:nowatermark_utility}
\setlength{\tabcolsep}{2.5pt}
\renewcommand{\arraystretch}{0.9}
\resizebox{\columnwidth}{!}{
\begin{tabular}{l*{12}{c}}
\toprule
\multirow{2}{*}{\textbf{Lang.}}
& \multicolumn{3}{c}{\textbf{B-HSJA}}
& \multicolumn{3}{c}{\textbf{DiffErase-L}}
& \multicolumn{3}{c}{\textbf{DiffErase-M}}
& \multicolumn{3}{c}{\textbf{ScrambleMark}} \\
\cmidrule(lr){2-4}
\cmidrule(lr){5-7}
\cmidrule(lr){8-10}
\cmidrule(lr){11-13}
& STOI & SPR & WER
& STOI & SPR & WER
& STOI & SPR & WER
& STOI & SPR & WER \\
\midrule
English
& 0.81 & 0.88 & 11.74\%
& 0.75 & 0.83 & 25.19\%
& 0.73 & 0.80 & 23.61\%
& 0.91 & 0.98 & 0.31\% \\

Chinese
& 0.83 & 0.85 & 13.96\%
& 0.72 & 0.83 & 19.90\%
& 0.74 & 0.79 & 21.38\%
& 0.92 & 0.96 & 0.72\% \\

Korean
& 0.82 & 0.85 & 26.81\%
& 0.71 & 0.81 & 27.63\%
& 0.70 & 0.76 & 23.27\%
& 0.91 & 0.99 & 0.53\% \\

French
& 0.85 & 0.87 & 14.36\%
& 0.72 & 0.83 & 23.04\%
& 0.71 & 0.79 & 19.91\%
& 0.93 & 0.97 & 0.49\% \\

German
& 0.84 & 0.82 & 18.62\%
& 0.70 & 0.85 & 20.01\%
& 0.73 & 0.80 & 24.55\%
& 0.92 & 1.00 & 1.05\% \\

\midrule
\textbf{Avg.}
& \textbf{0.83} & \textbf{0.85} & \textbf{17.10\%}
& \textbf{0.72} & \textbf{0.83} & \textbf{23.15\%}
& \textbf{0.72} & \textbf{0.79} & \textbf{22.54\%}
& \textbf{0.92} & \textbf{0.98} & \textbf{0.62\%} \\
\bottomrule
\end{tabular}}
\end{table}

\noindent\textbf{Results.}
ScrambleMark preserves clean-audio utility, achieving 0.92 STOI, 0.98 SPR, and 0.62\% WER on average, compared with 0.72--0.83 STOI, 0.79--0.85 SPR, and 17.10--23.15\% WER for the baselines (Table~\ref{tab:nowatermark_utility}). We also observe no false watermark detections on ScrambleMark-processed clean audio across all evaluated models.

\subsection{Human-Study Statistical Significance} \label{statisticalsignificance}

We test whether ScrambleMark's perceptual advantage over the baselines is statistically significant and consistent across participants. For each platform, eight participants each evaluate 120 clips (30 utterances $\times$ 4 methods); because participant groups differ across platforms, we analyze each platform separately and average the 30 utterance-level ratings within each participant and method to preserve the participant as the unit of analysis. Following~\cite{arias2021inexpensive}, we use the Friedman test~\cite{friedman1937use} for overall differences and paired Wilcoxon signed-rank tests~\cite{wilcoxon1945individual} to compare ScrambleMark with each baseline, with Holm correction~\cite{holm1979simple} for the three comparisons within each platform. We consider Holm-adjusted \(p<0.05\) statistically significant.

\noindent\textbf{Results.} The Friedman tests yield \(\chi^2(3)=24.00\) for both SynthID and TrustedAudio and \(\chi^2(3)=22.20\) for Resemble AI, with all \(p<0.001\). These results indicate highly consistent rank differences across participants on all three platforms. The pairwise Wilcoxon tests show \(W=0\) and raw \(p=0.0078\) across all comparisons between ScrambleMark and the baselines, indicating that, for each comparison, all participant-level mean differences favor ScrambleMark. After Holm correction within each platform, all comparisons remain significant (Holm-adjusted $p=0.0234<0.05$). Thus, the MOS advantages reported in Table~\ref{commercialperformance} are consistent across participants rather than driven by a small subset.

\section{Ethics Considerations}\label{sec:ethic}

\subsection{Vulnerability Disclosure}

We responsibly disclosed ScrambleMark to all affected vendors before public release, providing technical details, verification materials, and source code when requested. All vendors acknowledged our reports and had the opportunity to investigate and mitigate the issue. Google confirmed that the issue may be publicly disclosed.  


\subsection{Human Subjects Research}

Prior to the study, our protocol received institutional IRB approval. Participation was voluntary with informed consent, minimal risk, and no compensation, deception, or personally identifiable information (PII) collection; only anonymized ratings were retained.  Participants were recruited online from three universities in the same country and had sufficient English proficiency for the approximately 35-minute task.

\newpage 
\section{Meta-Review}

The following meta-review was prepared by the program committee for the 2027
IEEE Symposium on Security and Privacy (S\&P) as part of the review process as
detailed in the call for papers.

\subsection{Summary}

This paper presents ScrambleMark, a zero-query, watermark-agnostic black-box
evasion attack against speech watermarking systems. ScrambleMark disrupts the
harmonic structure of watermarked speech to evade detection and is evaluated
against both academic and commercial watermarking systems.

\subsection{Scientific Contributions}

\begin{itemize}
    \item Identifies an Impactful Vulnerability
    \item Creates a New Tool to Enable Future Science
    \item Provides a Valuable Step Forward in an Established Field
\end{itemize}

\subsection{Reasons for Acceptance}

\begin{enumerate}
    \item The paper identifies an impactful vulnerability in modern speech
    watermarking systems. The authors show that structural perturbations can
    evade watermark detection without querying the target detector during
    evasion, and demonstrate the attack against both academic and commercial
    systems.

    \item The paper provides a valuable step forward in an established field.
    ScrambleMark achieves high attack success while preserving substantially
    better speech quality than existing generative removal attacks.

    \item The paper creates a new tool to enable future science. ScrambleMark
    provides a concrete attack framework that can be used to study the
    robustness of speech watermarking systems against structural perturbations
    and to motivate the development of stronger defenses.
\end{enumerate}

\end{document}